\documentclass[12pt]{article}
\usepackage{graphicx}
\usepackage{amsmath}
\usepackage{amsfonts}
\begin{document}
\begin{center}
NUCLEAR PAIRING: NEW PERSPECTIVES\\[0.3cm]
Vladimir Zelevinsky$^{1,2}$ and Alexander Volya$^{3}$\\
{\small $^{1}$National Superconducting Cyclotron Laboratory and\\
$^{2}$Department of Physics and Astronomy,\\
Michigan State University, East Lansing, MI 48824-1321, USA,\\
$^{3}$Physics Division, Argonne National Laboratory, Argonne, IL 60439, USA}\\
\end{center}

\begin{abstract}
Nuclear pairing correlations are known to play an important role
in various single-particle and collective aspects of nuclear
structure. After the first idea by A. Bohr, B. Mottelson and D.
Pines on similarity of nuclear pairing to electron
superconductivity, S.T. Belyaev gave a thorough analysis of the
manifestations of pairing in complex nuclei. The current revival
of interest in nuclear pairing is connected to the shift of modern
nuclear physics towards nuclei far from stability; many loosely
bound nuclei are particle-stable only due to the pairing. The
theoretical methods borrowed from macroscopic superconductivity
turn out to be insufficient for finite systems as nuclei, in
particular for the cases of weak pairing and proximity of
continuum states. We suggest a simple numerical procedure of exact
solution of the nuclear pairing problem and discuss the physical
features of this complete solution. We show also how the continuum
states can be naturally included in the consideration bridging the
gap between the structure and reactions. The path from coherent
pairing to chaos and thermalization and perspectives of new
theoretical approaches based on the full solution of pairing are
discussed.

\end{abstract}

\section{Introduction}

Nuclear pairing is one of the main and longstanding pillars of
current understanding of nuclear structure. Pairing provides an
important contribution to the odd-even mass difference in the
phenomenological mass formulae \cite{BM}. As an empirical fact the
pairing was put in the foundation of the shell model by Mayer and
Jensen \cite{JM} in order to be able to predict ground state spins
and other properties of non-magic nuclei. In the shell-model
framework, the classification of paired states is usually
performed with the aid of the seniority scheme \cite{Rac, RT},
where the seniority counts a number of unpaired particles; a
similar scheme is used in atomic spectroscopy \cite{Judd}. The
Bardeen-Cooper-Schriffer (BCS) microscopic theory of
superconductivity \cite{BCS} elucidated the main features of the
ground state, excitation spectrum, transition probabilities and
phase transition in a Fermi-system governed by the attractive
pairing. Immediately after that A. Bohr, Mottelson and Pines
pointed out \cite{BMP} the similarity between the superconducting
pairing correlations and observed pairing effects in nuclei. The
thorough application of the BCS approach to the nuclear problem
was done by Belyaev in his seminal paper \cite{Bel}. It was
quantitatively demonstrated that the pairing correlations
influence nearly all phenomena in low-energy nuclear physics:
binding energy, single-particle spectra, transition probabilities,
collective vibrational modes, onset of deformation, rotational
moment of inertia, level density and thermal properties.

The BCS theory, as well as its advanced form the
Hartree-Fock-Bogoliubov (HFB) method \cite{HFB}, is formulated in
a way fully appropriate for macroscopic quantum systems; in fact
it gives an asymptotically exact solution \cite{Bog} in the
thermodynamic limit. For mesoscopic systems, as nuclei, atomic
clusters, quantum dots, Fullerenes, nanotubes or small metallic
grains, this approach, although qualitatively reflecting the main
physical features, turns out to be insufficient. The total number
$N$ of particles is preserved in this method only on average.
Since we have to describe the spectroscopy and reactions for a
specific nuclide, we need either to add special projection
procedures \cite{Lipkin60,Nogami64,Satula,sheikh00} that fix the
exact value of $N$ or generalize the formalism by approximately
including the matrix elements restoring the particle number
conservation \cite{Klein,Bang,cov,Bertsch}.

Another drawback of the BCS or HFB approaches is a sharp phase
transition as a function of parameters or temperature. As pointed
out by Belyaev \cite{Bel}, in a system with a discrete
single-particle spectrum, the Cooper phenomenon requires, in
contrast to a macroscopic Fermi-gas, a certain minimum strength of
pairing attraction. For a weaker pairing, the mean field
approaches, as BCS or HFB, give only a trivial normal solution,
while in reality the effects of pairing correlations still exist.
The pairing correlations in the mean field framework also vanish
immediately after the thermal phase transition. These predictions
are incorrect for mesoscopic systems. The exact shell model
calculations show \cite{big} that the pairing correlations do not
disappear at the BCS transition point revealing instead a long
tail of ``fluctuational superconductivity".

The main field of interest in nuclear structure is currently
shifted to the nuclei far from stability. As we move to loosely
bound systems, the influence of the continuum becomes exceedingly
important. Along with that, all attractive correlations are to be
taken into account properly in order to determine the position of
the drip line. The correct treatment of pairing as the main
attractive part of the residual interaction is absolutely
essential for such problems. Some nuclides, like the notorious
$^{11}$Li, are bound just due to the pairing correlations between
the outermost neutrons, an example of a real Cooper pair
\cite{Brog}. The theory of pairing including both discrete and
continuum single-particle levels is still in its infantry
\cite{Benn}.

Finally, there is a clear necessity to understand the interplay of
pairing with other parts of the residual nuclear interactions
going beyond the mean field approximation of the HFB method. Of
course, in the lower part of the nuclear chart ($p, sd$ and $pf$
shells) there are well developed modern shell model methods and
reliable effective interactions of the nucleons in the truncated
single-particle space, see for example
\cite{Poves,BW,richter91,wildenthal84}. With the possibility of
incorporating additional stochastic and statistical elements
\cite{SMMC,QMCD,ECM}, the shell model calculations are able to
describe an impressive amount of spectroscopic data.
Unfortunately, the qualitative interpretation of results obtained
by the large-scale shell-model diagonalization in terms of simple
physical models is getting quite difficult as the matrix
dimensions approach the limit of current computational strength.
In addition, one needs to mention that the continuum problem is
not solved in the standard shell model approach based on the
discrete spectrum. Therefore the gap between the shell model for
nuclear structure and the reaction theory is widening.

In this situation it is alluring to first separate the pairing
part of the nuclear interaction and to solve the corresponding
many-body problem exactly. As was shown in \cite{EP}, the exact
solution is numerically simple and eliminates all drawbacks
related to the BCS approximation. At the same time, it is still
close enough to the standard images of nuclear structure. This
exact solution can serve as a zero-order step or a background that
allows one to look for the new approaches and approximations to
the full problem, effects of other interactions, inclusion of the
continuum, relation to the reaction cross sections and so on.

In what follows we start with sketching the traditional approaches
and the exact solution of the pairing problem. We compare the
exact results with the BCS approximation, both for the ground and
excited states, demonstrate the possibility of including the
continuum physics and consider chaotic aspects of pairing, a topic
practically unexplored in the literature. We complete the paper
with the discussion of the perspectives of new approximations
based on the exact pairing solution.

\section{Approaching the solution of the pairing problem}

\subsection{Pairing Hamiltonian}

We formulate the pairing problem in the restricted single-particle
space of fermionic orbitals assuming the Hamiltonian
\begin{equation}
H_{p}=\sum_{1}\epsilon_{1} a^{\dagger}_{1} a_{1} \,-\,\frac{1}{4}
\sum_{1,2} G_{12}\,p^{\dagger}_1 p_{2}.
                                                   \label{BCS:ham}
\end{equation}
Here the subscripts {\sl 1}, ... run over the complete set of
orthogonal single-particle basis states, and we assume the Kramers
double-degeneracy of time-conjugate orbitals $|1)$ and
$|\tilde{1})$. The pair creation, $p^{\dagger}_{1}$, and
annihilation, $p_{1}$, operators are defined as
\begin{equation}
p^{\dagger}_{1}=a^{\dagger}_{1}a^{\dagger}_{\tilde{1}}, \quad
p_{1}=a_{\tilde{1}}a_{1},                   \label{p}
\end{equation}
and the double time-reversal acts as $|\tilde{\tilde{1}})=-|1)$.
In the important case of spherical symmetry of the mean field that
supports degenerate orbitals $|jm)$ with energies $\epsilon_{j}$,
angular momentum $j,$ and projection $j_{z}=m$, the pairing
Hamiltonian can be conveniently written as
\begin{equation}
H_p=\sum_j\,\epsilon_j \hat{N}_j - \sum_{j\,j'} \, G_{j j'}
L^\dagger_j L_{j'},                               \label{2.1}
\end{equation}
where we use the operator notations,
\begin{equation}
\hat{N}_{j}=\sum_{m}\hat{n}_{jm}=\sum_{m}a^{\dagger}_{jm}a_{jm},
                                                \label{2.2}
\end{equation}
for the occupancy operators and ($a_{\tilde{1}}\rightarrow
a_{j\tilde{m}}=(-)^{j-m}a_{j-m}$)
\begin{equation}
L_{j}=\frac{1}{2}\sum_{m}(-)^{j-m}a_{j-m}a_{jm}, \quad
L^{\dagger}_{j}=\frac{1}{2}\sum_{m}(-)^{j-m}
a^{\dagger}_{jm}a^{\dagger}_{j-m},             \label{2.3}
\end{equation}
for annihilation and creation operators of pairs in a state with
certain quantum numbers of total angular momentum and its
projection, $J=M=0$. Essentially the same Hamiltonian
(\ref{BCS:ham}) can describe the situation in the deformed mean
field.

The pairing interaction is defined in terms of real matrix
elements of diagonal pair attraction, $G_{11}>0$, and off-diagonal
ones, $G_{12}=G_{21}$, for pair transfer between the orbitals {\sl
1} and {\sl 2}. We limit ourselves here by the pairing of
identical particles (isospin $T=1$) in states with zero total pair
angular momentum; the consideration of the $T=0$ proton-neutron
pairing would require a non-zero spin of the pair.

\subsection{BCS approach}

In the BCS theory, the ground state $|0\rangle_{{\rm BCS}}$ of the
paired system with Hamiltonian (\ref{BCS:ham}) is determined by
minimizing the ground state energy with a trial  wave function
\begin{equation}
|0\rangle_{{\rm BCS}}=\Pi_{1> 0} (u_1 - v_1 p^{\dagger}_1) |0
\rangle=\Pi_{1> 0} u_1 \exp \left(-\frac{v_1}{u_1}\,p^\dagger_1
\right ) |0 \rangle\,, \label {BCS:wf}
\end{equation}
where the variational parameters for each pair of time-conjugate
orbitals, $u_{1}$ and $v_{1}$, can be taken as real numbers
subject to the normalization $u_{1}^{2}+v_{1}^{2}=1$. In our
discussion we denote $|N;s \dots \rangle$ as the lowest in energy
$N$-particle state with quantum numbers $s, \dots .$ We use the
notation $|s\rangle_{{\rm BCS}}$ for the BCS state with $s$
quasiparticles that has an uncertain particle number; for this
reason $N$ is not shown, however it is assumed that the state
corresponds to an average particle number $\overline{N}.$ The
restriction $1>0$ means that the time-conjugate orbitals are not
counted twice. The exponential form of the variational wave
function shows that this state is generated as a coherent state of
fermionic pairs; this feature can be put in a foundation of
methods going beyond the BCS \cite{rowe95,yi91}.

The variational solution is given via the occupation amplitudes
\begin{equation}
v_1^2=n_1=\frac{1}{2}\left(1-\frac{\epsilon_1}{e_1}\right),\quad
u_1^2=1-n_1=\frac{1}{2}\left(1+\frac{\epsilon_1}{e_1}\right),
                                         \label{BCS:uandv}
\end{equation}
where $e_1=\sqrt{\epsilon_1^2+\Delta_1^2} $ is quasiparticle
energy  and $\Delta_1$ is the BCS energy gap. The gap equation
arising from the minimization of energy is
\begin{equation}
\Delta_1=\frac{1}{2}\, _{{\rm BCS}}\langle 0|\sum_2
G_{1\,2}\,p^{\dagger}_2 |0\rangle_{{\rm BCS}}=\sum_2
\frac{G_{1\,2}}{2e_2}\,\Delta_2.
                                             \label{BCS:gap}
\end{equation}
The ground state as well as low-lying excited states of paired
systems can be classified introducing quasiparticle creation
operators
\begin{equation}
\alpha^{\dagger}_1=u_1 a^{\dagger}_1- v_1 a_{\tilde{1}}\,.
                                             \label{bog}
\end{equation}
This transformation is canonical due to the correct normalization
of $u$ and $v$. The BCS vacuum $|0\rangle_{{\rm BCS}}$ in
(\ref{BCS:wf}) can be defined as $\alpha |0\rangle_{{\rm
BCS}}=0\,.$ The Bogoliubov transformation (\ref{bog}) mixes
particle and hole states.

One of the problems in the BCS application to small systems is
particle number non-conservation which follows from the form of
the wave function (\ref{BCS:wf}). A common practice to fix the
number of particles is to introduce a chemical potential $\mu$ by
the shift of single-particle energies $\epsilon\rightarrow
\epsilon-\mu.$ The right total particle number is restored in
average through the condition
\begin{equation}
\sum_1 n_1(\mu) = \overline{N}.                       \label{mu}
\end{equation}
The uncertainty of the total particle number is given by
\begin{equation}
(\Delta N^2) =\overline{N^2}-{\overline{N}}^2= \sum_1
n_1(1-n_1)\,.                               \label{BCS:unc}
\end{equation}
Practically this fluctuation is of the order $\sqrt{(\Delta
N^2)}\approx 2\,.$ Given that relative fluctuations go down with
increasing $N\,,$ the BCS solution is asymptotically exact in the
thermodynamic limit of macroscopic systems \cite{Bog}. The
contribution to the energy from the fluctuation is quadratic in
the number of particles \cite{rowe} and thus gives an extra
correction of the monopole type. Special methods, such as
Lipkin-Nogami \cite{Lipkin60,Nogami64} techniques, were invented
in order to suppress the particle number fluctuations.

The particle number violation in the BCS is an example of
spontaneous symmetry  breaking with respect to phase rotations
generated by the number operator, $U(\phi)=\exp (-i \phi
\hat{N}/2)\,. $ While the pairing Hamiltonian is invariant under
this rotation, $[H_p,\,U(\phi)]=0\,,$ the trial ground state
$|0\rangle_{{\rm BCS}}$  has a preferred orientation, with the usual
consequences of the appearance of Anderson-Goldstone-Nambu modes
whose properties are influenced by the mesoscopic nature of the
nuclear systems \cite{broglia00}. A number of projection
techniques have been developed within the framework of the
symmetry violation treatment, see \cite{RingSchuck} and references
therein.

\subsection{Recursive method with particle number conservation}

Instead of introducing the pair condensate of the pairs with an
uncertain particle number, another possibility was explored
\cite{dang66,dang68,cov}, where the matrix elements of relevant
operators explicitly keep memory of the exact particle number. The
gap is defined now as a matrix element of the pair annihilation
operator between the ground states $|N;0\rangle$ of the
neighboring even systems,
\begin{equation}
\Delta_{1\,}(N)=\frac{1}{2}\,\langle N-2;0| \sum_{2}\,G_{1\,2} \,
p_{2}  |N;0\rangle \,.                          \label{PC:gap}
\end{equation}
Similarly, the particle number dependence enters the
single-particle transition amplitudes between adjacent even and
odd systems,
\begin{equation}
v_{1\,}(N)=\langle N-1; \tilde{1\,}|a_{1\,}|N;0\rangle,\, \quad
u_{1\,}(N)=\langle N+1; 1\,|a^{\dagger}_{1\,} |N;0 \rangle \,.
                                                  \label{PC:uv}
\end{equation}
Here one needs to consider a sequence of ground states
$|N;0\rangle$ with energies $E(N)\,$. It is assumed that the
spectra of adjacent odd nuclei start with energies $E(N\pm
1;\,1\,)$ of the states $|N\pm1;1\,\rangle$ containing one
unpaired nucleon with quantum numbers {\sl 1}.

The exact operator equations of motion for the single-particle
operators $a_1\,$ and $a^\dagger_1\,$,
\begin{equation}
[a_1\,,\,H]=\epsilon_1\, a_1\,\,+\, \frac{1}{2} \sum_{2}\,
G_{1\,2}\, a^{\dagger}_{\tilde{1\,}}\, p_{2}\,,
                                        \label{PC:motion1}
\end{equation}
\begin{equation}
[a^{\dagger}_{1\,},\,H]=-\epsilon_1\, a_1^{\dagger}\,-\, \frac{1}{
2}\sum_{2}\, G_{1\,2}\, p^{\dagger}_{2}\,a_{\tilde{1\,}},
                                           \label{PC:motion2}
\end{equation}
can be used to construct recursive in $N$ equations for the gap
(\ref{PC:gap}) and single-particle transition amplitudes
(\ref{PC:uv}). The approximation of no condensate disturbance by
an extra particle \cite{dang66,dang68},
\[\langle N-1; 1\, |\sum_{2}\,G_{1\,2}\, a^{\dagger}_{\tilde{1\,}}\,
p_{2}|N;0\rangle\]
\begin{equation}
\approx \langle N-1; 1\,|a^{\dagger}_{\tilde{1\,}} |N-2;0\rangle
\langle N-2;0| \sum_{2}\,G_{1\,2}\,\, p_{2}|N;0\rangle =2
\Delta_{1\,}(N)\,u_{\tilde{1\,}}(N-2),     \label{PC:appr}
\end{equation}
leads to the recursion relation connecting adjacent even nuclei,
\begin{equation}
\left |v_{1\,}(N-2) \right |^2=1- \frac{\left | \Delta_{1\,}(N)
\right |^2}{[e_{1\,}(N)-{\epsilon'}_{1\,}(N)]^2}\left |v_{1\,}(N)
\right |^2\,,                             \label{PC:recursionv}
\end{equation}
where the $N$-dependent chemical potential is introduced,
\begin{equation}
\mu(N)=\frac{1}{2}\left (E(N)-E(N-2) \right ),    \label{PC:mu}
\end{equation}
and quasiparticle excitation energy is defined as
\begin{equation}
e^2_{1\,}={{\epsilon'_1(N)}^2 +|\Delta_{1\,}(N)|^2} \,,
                                                    \label{7}
\end{equation}
with shifted single-particle energies
\begin{equation}
\epsilon_1'(N)=\epsilon_{1\,}-\frac{G_{1\,1\,}}{2} -\mu(N)\,.
                                            \label{eprime}
\end{equation}
The analogs of the number conservation equation and the gap
self-consistency condition now read
\begin{equation}
\Omega-N+2=\sum_{1\,}\frac{\left | \Delta_{1\,}(N)\right |^2}
{[e_{1\,}(N)-{\epsilon'}_{1\,}(N)]^2}\left |v_{1\,}(N) \right |^2,
                                             \label{PC:chem}
\end{equation}
where $\Omega$ is the total capacity of fermionic space, and
\begin{equation}
\Delta_{1\,}(N)=
\frac{1}{2}\,\sum_{2} \, G_{1\,2} \frac{\Delta_{2}(N)\, \left
|v_{2}(N) \right |^2}{e_{2}(N)-{\epsilon'}_{2}(N)} \,.
                                             \label{PC:gap1}
\end{equation}

The pairing problem formulated in this manner allows a recursive
solution in both directions, starting from the empty shell or from
the completely filled shell. This solution reduces to the BCS
under assumption that the gap does not change in the transition
from $N$ to $N-2,$ the same approximation of particle number
uncertainty that lead to the BCS particle number fluctuation
(\ref{BCS:unc}). Based on this feature, the BCS energy can be
efficiently corrected by the substitution $\overline{N}\rightarrow
\overline{N}-1$ \cite{PC}. Corrections to such iterative methods
via inclusion of pair-vibration excitations in the intermediate
states of Eq. (\ref{PC:appr}) with further diagonalization are
also possible \cite{cov}, as well as the treatment of the
excitations with RPA techniques \cite{hogaasen61,johns70}.

The particle conserving treatment does not resolve another problem
of the BCS solution, namely the sudden disappearance of pairing
correlations when coupling becomes too weak. The gap equations,
(\ref{BCS:gap}) and (\ref{PC:gap1}), have only trivial $\Delta=0$
solutions if the pairing strength $G$ is too small compared to the
single-particle energy spacings. The point of this phase
transition is roughly at the critical coupling strength $G_{c}$,
\begin{equation}
G_{c}\, \nu_{\rm F} = 1,                       \label{disapp}
\end{equation}
where both the pairing strength and the density of single-particle
states $\nu_{\rm F}$ are taken at Fermi energy. Many nuclear
systems in the shell model picture are close to or even below the
point of the BCS instability, although the pairing correlations do
still exist \cite{EP}. As will be discussed later, near the phase
transition, in the so-called pair-vibrational regime, the
fluctuations drive pair scattering to an almost chaotic level
leading to a sharp increase of the mixing between the states of
the same seniority. This randomness makes the approximation
(\ref{PC:appr}) or any truncation of states mixed by the pair
vibrations inappropriate. Various projection techniques also
seriously suffer in the region of weak pairing. More advanced
approaches, such as HFB+RPA, break down in the vicinity of the
phase transition, though the pairing solution can still be
continued into the region beyond the critical point using the RPA
based on the Hartree-Fock solution for a normal state. This
treatment drastically improves the prediction for the ground state
energy \cite{hagino00}. The methods of equations of motion
\cite{rowe,rowe68} and variational techniques can be used to
better account for the RPA ground state correlations
\cite{dukelsky90}. Being applied to superfluid Fermi systems these
methods demonstrated a considerable improvement
\cite{dukelsky90,passos98}.

\section{Exact solution of the pairing problem}

Historically, a few suggestions were put forward for the exact
solution of the pairing problem. The Richardson method, described
in the series of papers \cite{Rich,richardson64}, provides a
formally exact way for solving the pairing Hamiltonian. This
method reduces the large-scale diagonalization of a many-body
Hamiltonian matrix in a truncated Hilbert space to a set of
coupled equations ($\Omega_{j}=2j+1$)
\begin{equation}
\sum_j \frac{\Omega_j}{2\epsilon_j-z_\lambda}-\sum_{\lambda'\ne
\lambda}\, \frac{4}{z_{\lambda'}-z_{\lambda}}=\frac{2}G\,
                                               \label{rich}
\end{equation}
for unknown parameters $z_\lambda\,,$ their number being equal to
that of valence particle pairs. The ground state energy is then
equal to $E(N)=\sum_{\lambda}^{N/2}\,z_\lambda\,.$

Recently this solution was revived and reinterpreted \cite{Pitt}
with the aid of the electrostatic analogy, similar to that used by
Dyson in his theory \cite{Dyson} of random level ensembles.
Unfortunately, the Richardson solution is only valid for the
constant pairing force, $G_{jj'}=G=\,$const. It also requires
serious numerical efforts rapidly growing with the number of
particles. Recently, exact solutions have been also approached
with sophisticated mathematical tools as infinite-dimensional
algebras \cite{pan98}. Such formally exact solutions have a
certain merit from a mathematical point of view and might be
useful for developing simple models \cite{pan00,dukelsky00}.
However, they are not very promising for practical problems in
nuclear physics.

The natural way of solving the pairing problem is related to the
direct Fock-space diagonalization. For deformed nuclei with the
doubly degenerate single-particle orbitals this approach
supplemented by the appropriate use of symmetries and truncations
was already shown to be quite effective
\cite{burglin96,molique97}. The diagonalization of the general
pairing Hamiltonian (\ref{2.1}) is much simpler than that of the
full shell-model Hamiltonian due to the possibility of classifying
many-body states within the seniority scheme \cite{Rac,RT,Kerman},
especially in the case of spherical symmetry (\ref{2.1}). Long ago
it was shown \cite{Kerman,Auer} that this approach is useful not
only in the exactly solvable degenerate model but in a realistic
shell model context as well. With a perspective to complement the
pairing problem with the subsequent account for other parts of the
residual interaction, we consider this path promising and quite
practical.

It is well known that the pair annihilation, $L_{j}$, pair
creation $L^{\dagger}_{j}$, and occupation number operator
(shifted to the middle of the $j$-subshell),
\begin{equation}
L_{j}^{\circ}=\frac{1}{2}\hat{N}_{j}-\frac{1}{4}\Omega_{j}, \quad
\Omega_{j}=2j+1,                                  \label{2.4}
\end{equation}
form an SU(2) algebra of ``quasispin" for each $j$-subshell,
\begin{equation}
[L_{j},L^{\circ}_{j'}]=\delta_{jj'}L_{j}, \quad [L^{\dagger}_{j},
L^{\circ}_{j'}]=-\delta_{jj'}L^{\dagger}_{j}, \quad [L_{j},
L^{\dagger}_{j'}]=-2\delta_{jj'}L^{\circ}_{j}.
                                                    \label{2.5}
\end{equation}
Therefore the pairing Hamiltonian (\ref{2.1}) preserves all
partial quasispins $\Lambda_{j}$,
\begin{equation}
{\bf L}_{j}^{2}=(L^{\circ}_{j})^2+\frac{1}{2}(L_{j}^{\dagger}L_{j}
+L_{j}L^{\dagger}_{j})=\Lambda_{j}(\Lambda_{j}+1).
                                                   \label{2.6}
\end{equation}
The partial seniority quantum numbers,
\begin{equation}
s_{j}=\frac{\Omega}{2}-2\Lambda_{j},                    \label{2.7}
\end{equation}
are also conserved. They express the number of unpaired, and
therefore not participating in the pairing interaction
(\ref{2.1}), particles. The fully paired $j$-level corresponds to
the maximum partial quasispin $\Lambda_{j}= \Omega/4$ and lowest
partial seniority $s_{j}=0$.

The pair transfer $L^{\dagger}_{j'}L_{j}$ between the levels
$j\rightarrow j'$ changes the occupancies, i.e. projections
$L^{\circ}_{j}$ and $L^{\circ}_{j'}$, keeping intact the lengths
of quasispins $\Lambda_{j}$ and $\Lambda_{j'}$, and, whence,
seniorities $s_{j}$ and $s_{j'}$. The space is decomposed into
sectors with given partial seniorities $s_{j}$, and the basis
states within each sector can be labelled by the set of
occupancies $N_{j}$ under a constraint $\sum_{j}N_{j}=N$, the
total valence particle number. The passive (unpaired) particles
occupy fixed orbitals and create nonzero seniorities. They
influence the dynamics indirectly, through the Pauli blocking. The
states with zero total seniority $s=\sum_{j}s_{j}$ have the total
spin $J=0$, while for $s\neq 0$ the further decomposition with
respect to the rotation group is possible, and some many-body
states with different angular momentum coupling but the same
seniorities remain degenerate.

Using the states with given values of $s_{j}$ and various possible
occupancies $N_{j}$ as a basis, it is easy to construct the
Hamiltonian matrix that is essentially the matrix with respect to
the sets of $N_{j}$. The diagonal matrix elements are
\begin{equation}
\langle\{s_{j}\},\{N_{j}\}|H_{p}|\{s_{j}\},\{N_{j}\}\rangle=
\sum_{j}\left[\epsilon_{j}N_{j}-\frac{G_{jj}}{4}(N_{j}-s_{j})
(\Omega_{j}-s_{j}-N_{j}+2)\right].                 \label{2.8}
\end{equation}
Each term in the square brackets gives a full solution for the
pairing problem on a degenerate $j$-level. Clearly, as long as
seniority is small, $s_{j}\ll \Omega_{j}$, each unpaired particle
increases energy by $\Delta_{j}=G_{jj}\Omega_{j}/4$, and this
quantity plays the role analogous to that of the energy gap in the
BCS theory. The off-diagonal matrix elements for the pair transfer
$j'\rightarrow j$ are
\begin{equation}
H_{j'\rightarrow j}=-\frac{G_{jj'}}{4}\left[(N_{j'}-s_{j'})
(\Omega_{j'}-s_{j'}-N_{j'}+2)(\Omega_{j}-s_{j}-N_{j})(N_{j}-s_{j}+2)
\right]^{1/2}.                                  \label{2.9}
\end{equation}

The highest matrix dimension is encountered for the lowest
possible total seniority, $s=0$ for an even number of particles,
and $s=1$ for an odd number of particles. But, even for heavy
nuclei, this dimension does not exceed few thousand (in modern
shell model computations one has to deal with dimensions $10^{8}$
and higher in the $m$-scheme). In addition, the Hamiltonian matrix
is very sparse. As a result of the numerical diagonalization we
obtain the spectrum of states for a given set of seniorities. For
example, for an even system, the condition $s=0$ selects all zero
partial seniorities, $s_{j}=0$. All those states correspond to
pair condensates that differ by the distribution of the average
partial occupancies $\{ N_{j}\}$ among the subshells. In a
standard language of the BCS theory supplemented by the random
phase approximation (RPA), the excited states for $s=0$ are
various pair vibrations. However, here we do not make any
assumptions of boson character or harmonic spectrum of
excitations. The next section illustrates the typical results of
the diagonalization.

\section{Example: A chain of even isotopes}

The longest known chain of tin isotopes is a subject of extensive
experimental and theoretical studies. Even considering the proton
subsystem, $Z=50$, as an inert core, we have to deal with the
neutron model space that is too large for a direct
diagonalization. Modern computational techniques that use the
Lanczos iteration method allow for exact determination of few
low-lying states in systems with up to 12 valence particles
\cite{holt98}. These results are essential for testing the
approximate techniques. It is known that pairing correlations play
a major role in forming the ground state wave functions of tin
isotopes.

\begin{table}
$$
\begin{array}{c|ccccc}
& g_{7/2} & d_{5/2} & d_{3/2} & s_{1/2} & h_{11/2} \cr
\epsilon_j &-6.121& -5.508 & -3.749 & -3.891 & -3.778 \cr
\hline
g_{7/2}&0.9850 & 0.5711 & 0.5184 & 0.2920& 1.1454 \cr
d_{5/2}& & 0.7063 & 0.9056 & 0.3456 &  0.9546 \cr
d_{3/2}& & & 0.4063 & 0.3515 & 0.6102 \cr
s_{1/2}& & & & 0.7244 & 0.4265 \cr
h_{11/2}& & & & & 1.0599 \cr
\end{array}
$$
\caption{Single-particle energies and pairing matrix elements
$V_{0}(jj;j'j')$ (in MeV) for the shell model space from
$^{100}$Sn to $^{132}$Sn; matrix elements are determined from
$G$-matrix calculations. \label{Gpair}}
\end{table}

Unlike in many other nuclear systems, pairing in tin isotopes is
quite strong and stable being sufficiently above the point of the
BCS phase transition. There is only a relatively minor weakening
in the mass region near $^{114}$Sn due to a gap between $d_{5/2}$
and $g_{7/2}$ and the rest of single-particle orbitals. We
specifically would like to explore this region in order to discuss
physics of the BCS phase transition.

For tin isotopes ranging in mass number from $A=100$ to $A=132$,
we assume a configuration space between the two neutron magic
numbers, 50 and 82. The valence neutron space contains here five
single-particle orbitals,
$h_{11/2},\,d_{3/2},\,s_{1/2},\,g_{7/2},$ and $d_{5/2}\,.$ We
adopt parameters shown in Table \ref{Gpair}, the single-particle
energies taken from experimental data and the interaction matrix
elements from the $G$-matrix calculation \cite{holt98}. The
interaction parameters $V_{0}$ in Table 1 are related to the
pairing strengths $G_{j j'}$ as
\begin{equation}
G_{jj'}=\frac{1}{2}V_0(jj;j'j')\sqrt{(2j+1)(2j'+1)}.
                                               \label{V0}
\end{equation}
The shell model calculations with these parameters reproduce the
spectroscopy of tin isotopes in the region $A=120$ to $A=130$ with
a good accuracy. In parallel we discuss similar effects in calcium
isotopes where we used a well established FPD6 interaction
\cite{richter91}. The $fp$ neutron valence space covers calcium
isotopes from $^{40}$Ca to $^{60}$Ca. The weakening of pairing in
Ca occurs near $^{48}$Ca, i.e. for the $f_{7/2}$ subshell closure.
Results of the calculation for the $^{114}$Sn region are shown in
Table \ref{sntable} and for the region $^{48}$Ca in Table
\ref{catable}.

\begin{table}
$$
\begin{array}{c|c|ccccc}
\hline && g_{7/2} & d_{5/2} & d_{3/2} & s_{1/2} & h_{11/2} \cr
\hline \hline &EP: &&& &&\cr \hline

({\rm a})&N_j(N)        &6.96&4.46& 0.627& 0.356& 1.6\cr ({\rm
b})&n_j(N)        & 0.870 &0.744& 0.157 &0.178& 0.133\cr

({\rm c})&u^2_j(N)      &  0.128& 0.252& 0.838& 0.817& 0.863\cr
({\rm d})&v^2_j(N)      &  0.865& 0.736 &0.155& 0.177& 0.131\cr
({\rm e})&S_j(N+1)     &  2.8& 3.13 &3.14 &3.39& 3.29\cr ({\rm
f})&S_j(N)       &  6.86 &6.55 &7.25& 6.98& 7.12\cr ({\rm
g})&|\langle N+2;0 |P^\dagger_j|N;0\rangle|  & 0.68 &0.779 &0.617&
0.514& 1.03\cr ({\rm h})&|\langle N;0|P_j|N-2;0\rangle|  & 0.81
&0.93& 0.524& 0.396& 0.845\cr \hline \hline
&BCS: & &  &&& \cr
\hline

({\rm i})&N_j(N)      & 6.71 &4.14& 0.726 &0.507& 1.91\cr ({\rm
j})&n_j(N)      &    0.839 &0.69 &0.181& 0.254& 0.159\cr ({\rm
k})&\Delta_j     &     1.31 &1.43& 1.43& 1.38& 1.25\cr ({\rm
l})&e_j    &    1.78& 1.55 &1.86& 1.59& 1.71\cr ({\rm m})&S_j(N+1)
&     2.89& 3.21& 3.11 &3.21& 3.26\cr ({\rm n})&S_j(N)     & 6.89
&6.64& 7.2& 7.03 &7.06\cr ({\rm o})&_{{\rm BCS}}\langle
0|P_j|0\rangle_{{\rm BCS}}  & 0.734& 0.801& 0.545& 0.435 &0.896\cr
\hline
\end{array}
$$
EP: $E(^{114}{\rm Sn})=-86.308,\; E(^{116}{\rm Sn})=-95.942,\;
E(^{112}{\rm Sn})=-75.831.$ \\
BCS: $E(^{114}{\rm Sn})= -85.938,\;\mu= -5.035$.

\caption{The results of the exact pairing solution (EP) compared
to the BCS solution for the $^{114}$Sn nucleus; the interaction
matrix elements are determined by the $G$-matrix calculations, see
text; the separation energies $S_{j}$, quasiparticle energies
$e_{j}$ and the pairing gaps $\Delta_{j}$ are given in MeV.
\label{sntable}}
\end{table}

An important consequence of the proximity to the BCS phase
transition is a reduction of pairing correlation energy predicted
by the BCS but not confirmed by the exact solution. In the tin
example the BCS underpredicts binding energy by about 0.4 MeV,
while for calcium this number reaches 0.6 MeV. A similar
difference appears in one-nucleon separation energy
$S_j(N)=E(N-1)-E(N)$ (index $j$ here denotes the orbital of the
unpaired nucleon in an odd-$N$ system), as can be seen from the
comparison of rows (e) and (f) with the corresponding BCS
prediction, lines (m) and (n), Tables 2 and 3. As stressed in Ref.
\cite{EP}, this discrepancy can be crucial for the nuclei near
drip lines. Unfortunately, the BCS hardly can be improved with
respect to the treatment of weak pairing. Even for the complicated
particle number projection techniques accompanied by the the
variational procedures on a broader set of mean field states, it
remains unclear to what extent it is possible to describe the
pairing phase transition, and whether the high-lying pair
vibrations are included, the step needed to account for missing
correlation energy.

\begin{table}
$$
\begin{array}{c|c|cccc}
\hline &j&7/2&3/2&5/2&1/2 \cr &\epsilon_j&-8.39&-6.5&-1.9&-4.48
\cr \hline \hline &EP: &&&&\cr \hline ({\rm a})&N_j(N)        &
6.87&0.85&0.173&0.111\cr ({\rm b})&n_j(N)        &
0.858&0.212&0.0288&0.0557\cr ({\rm c})&u^2_j(N)      &
0.133&0.779&0.97&0.939\cr ({\rm d})&v^2_j(N)      &
0.848&0.212&0.0281&0.0555\cr ({\rm e})&S_j(N+1)     &
4.48&5.78&1.55&4.09\cr ({\rm f})&S_j(N)       &
9.64&9.75&14.1&11.5\cr ({\rm g})&\langle
N+2;0|P^\dagger_j|N;0\rangle & 0.706 &0.928&0.289&0.309 \cr ({\rm
h})&\langle N;0|P_j|N-2;0\rangle & 1.07&0.612&0.288&0.232 \cr
\hline \hline &BCS: & & && \cr \hline ({\rm i})&N_j(N)      &
6.5&1.22&0.155&0.124\cr ({\rm j})&n_j(N)      &
0.813&0.304&0.0258&0.062\cr ({\rm k})&\Delta_j     &
1.66&1.44&1.73&1.53 \cr ({\rm l})&e_j    & 2.13&1.56&5.45&3.18\cr
({\rm m})&S_j(N+1)   &    4.5&5.66&1.64&4.1 \cr ({\rm n})&S_j(N) &
9.64&9.51&13.8&11.3 \cr ({\rm o})&_{{\rm BCS}}\langle
0|P_j|0\rangle_{{\rm BCS}} & 0.78&0.651&0.275&0.241 \cr \hline
\end{array}
$$
EP: $E(^{48}{\rm Ca})=-71.215,\; E(^{50}{\rm Ca})=-85.149,\; E(^{46}{\rm Ca})=-55.501$.\\
BCS: $E(^{48}{\rm Ca})= -70.591, \;\mu= -7.335.$

\caption{Properties of the weakly-paired $^{48}$Ca nucleus; the
FPD6 interaction  was used in these calculations; all energies are
given in MeV. \label{catable}}
\end{table}

Another related feature is the difference in predicted occupation
numbers that can be inferred from comparing rows (a) and (b) with
(i) and (j) in Tables 2 and 3. A proper account for this
difference can partially help to correct the binding energy. In
the presence of additional interactions, the monopole contribution
to the energy can be particularly sensitive to the precise
occupation numbers. Furthermore, in the use of mean-field methods
for paired systems, a good reconstruction of the density matrix
generated by the pairing is of critical importance.

The exact pairing treatment (EP) becomes increasingly important in
considering the reaction amplitudes with paired nuclei. The
one-nucleon transition amplitudes defined in the exact solution
via Eq. (\ref{PC:uv}) can be compared with the corresponding BCS
quantities. Since, similarly to the recursive approach, these
amplitudes connect different nuclei, the standard BCS relations
$v^2_j=n_j$, $u_j^2=1-n_j$ and $u_j^2+v^2_j=1$ are no longer true.
Deviations from these equalities are clearly enhanced in the phase
transition region where adding an extra particle can make a sharp
difference. The BCS theory with an uncertain particle number does
not account for such effects. The pair emission amplitudes exhibit
even larger differences. Rows (g) and (h) of the Tables 2 and 3
show these amplitudes for adjacent even systems. The numbers are
noticeably different (in the BCS approach they are substituted
with a single set shown in row (o)). These discrepancies are
particularly crucial for weakly bound nuclei since not only the
binding energy is affected by the improved treatment of pairing
but also there are significant corrections to the reaction
amplitudes.

\begin{figure}
\begin{center}
\includegraphics[width=14 cm]{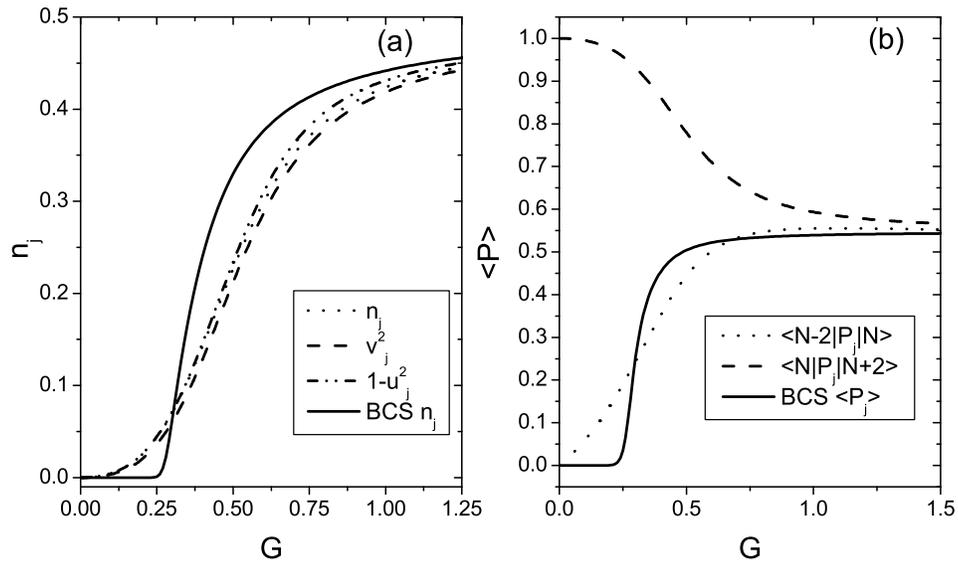}
\end{center}
\caption{Comparison of BCS and EP solutions for the ladder system
as a function of pairing strength $G$. In panel (a) occupation
numbers for the sixth level (first level above the Fermi surface)
and spectroscopic factors for capture and decay are compared to
exact calculation. For the same level, the pair emission and pair
absorption amplitudes are compared, panel (b), to the BCS
prediction. \label{ladepbcs}}
\end{figure}

Further insight into the situation can be gained by varying the
coupling strength. For this purpose we consider a ``ladder'' model
that contains ten double-degenerate single-particle orbitals
equally spaced with the interval of a unit of energy. The valence
space is assumed to be half-occupied with $N=10$ particles. The
most interesting region is near the Fermi surface, $\epsilon_{\rm
F}=5.$ For Fig. \ref{ladepbcs} we consider the first
single-particle level above the Fermi energy. As in the previous
example, the BCS result significantly deviates from the exact
solution near the phase transition, around $G=0.5$, as seen from
panel (a). In the same region one can observe a slight difference
between  $v_j^2$, $1-u_j^2$ and $n_j$ in the exact solution. For
the pair emission process, the differences between the BCS and
exact solution become more pronounced. Here the particle number
uncertainty is crucial since the level under consideration is
above Fermi energy for $N=10$, but below it for $N=12$.

\begin{figure}
\begin{center}
\includegraphics[width=14 cm]{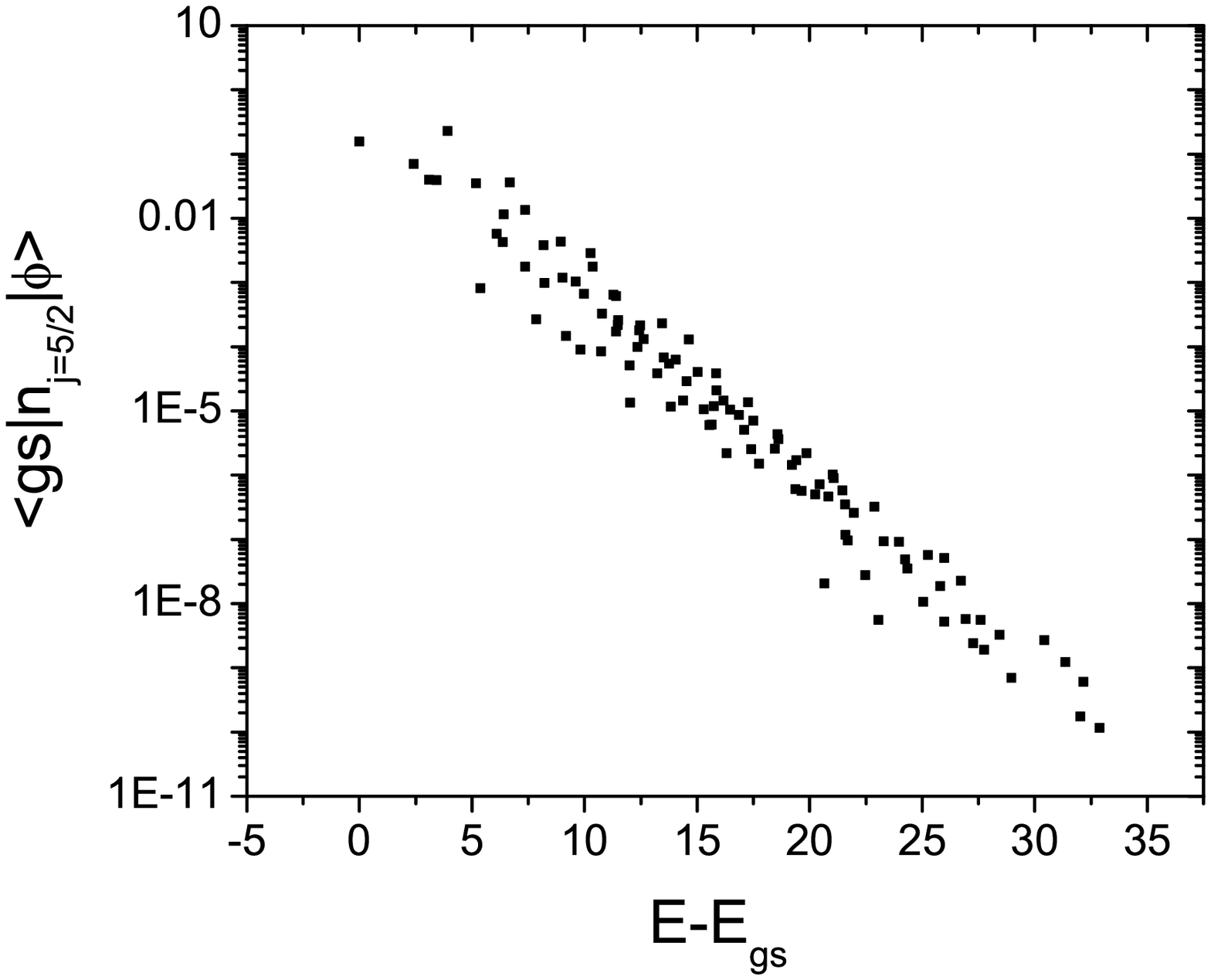}
\end{center}
\caption{Matrix elements of the operator $N_j$ for the $d_{5/2}$
level in $^{114}$Sn between the ground state and all pair
vibrational states $\phi$ (zero seniority) plotted as a function
of excitation energy of the state $\phi$. \label{sn_n}}
\end{figure}

\begin{figure}
\begin{center}
\includegraphics[width=14 cm]{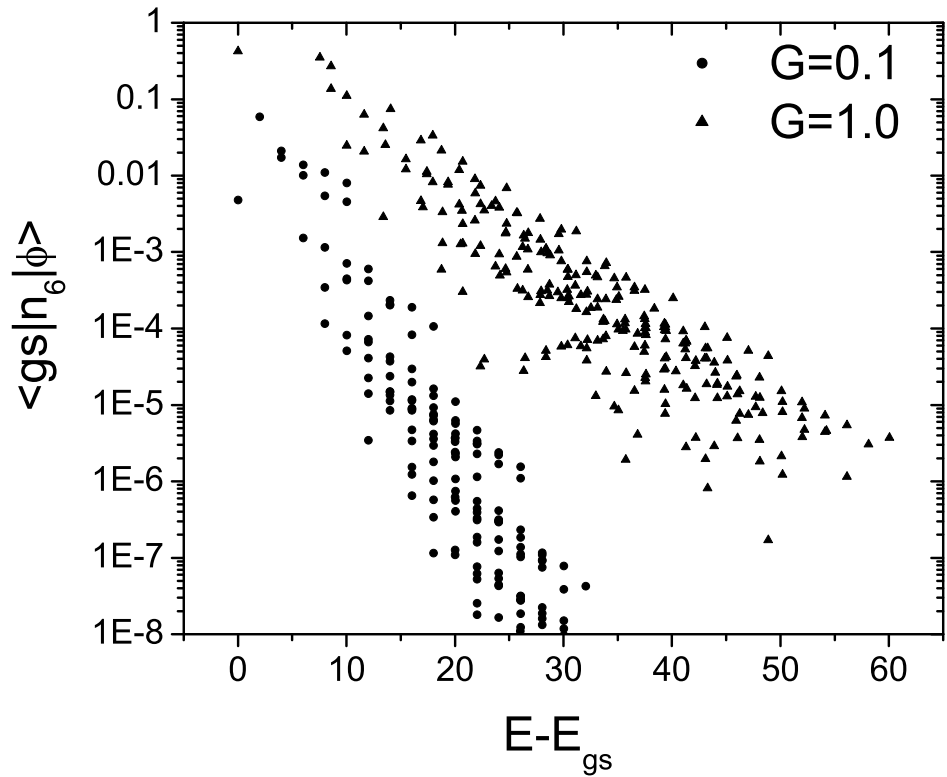}
\end{center}
\caption{Matrix elements of $N_6$ (the sixth single-particle state
in the ladder model) between the ground state and excited $s=0$
states as a function of excitation energy. \label{ladder_n}}
\end{figure}

Contrasting the exact solution with the mean field picture, we can
notice that the occupation number operators $\hat{N}_j$ in general
may have nonzero off-diagonal matrix elements between states of
the same seniority. In Figs. \ref{sn_n}, $^{114}$Sn, and
\ref{ladder_n}, the ladder model, the matrix elements between the
ground state and all $s=0$ states are shown as a function of
excitation energy. In all cases the off-diagonal matrix elements
rapidly fall off. In the case of weak pairing (Fig.
\ref{ladder_n}, circles), one can still see the structure of
excited states based on the equidistant single-particle spectrum.
For stronger pairing (compared to the single-particle level
spacing), Fig. \ref{ladder_n}, triangles, as well as in the
realistic case for spherical symmetry, Fig. \ref{sn_n}, the
decrease of matrix elements is more uniform and can be
approximated in average by an exponential function of excitation
energy. This indicates chaotization of motion even in the sector
with seniority $s=0$, \cite{vran}. Only very few states with
relatively large matrix elements may carry pair-vibrational
features. As follows from the extended shell model analysis
\cite{fraz}, the exponential tails of the strength functions are
typical for many-body quantum chaos \cite{gur}. The property of
exponential convergence was demonstrated in Ref. \cite{ECM} and
the extrapolation based on this property was later used
\cite{hor,cov1,print} as a practical tool for getting reliable
quantitative results in shell-model calculations of intractable
large dimensions.

\section{Continuum effects}

As the main interest of low-energy nuclear physics is moving to
nuclei far from the valley of stability, the continuum effects
become exceedingly important for a unified description of the
structure of barely stable nuclei and corresponding nuclear
reactions. The pairing part of the residual interaction in some
cases is the main source of the nuclear binding; spectroscopic
factors and reaction amplitudes are also critically dependent on
pairing.

As a demonstration of a realistic shell-model calculation
combining the discrete spectrum and the continuum, we consider
oxygen isotopes in the mass region $A=16$ to 28. In this study we
use a universal $sd$-shell model description with the
semi-empirical effective interaction (USD) \cite{wildenthal84}.
The model space includes three single-particle orbitals
$1s_{1/2}$, $0d_{5/2}$ and $0d_{3/2}$ with corresponding
single-particle energies $-3.16,\; -3.95$ and 1.65 MeV. The
residual interaction is defined in the most general form with the
aid of a set of 63 reduced two-body matrix elements in pair
channels with angular momentum $L$ and isospin $T$,
$\langle(j_{3}\tau_{3},j_{4}\tau_{4})LT|V|(j_{1}\tau_{1},
j_{2}\tau_{2})LT\rangle$, that scale with nuclear mass as
$(A/18)^{-0.3}$.

In the discrete spectrum the full shell model treatment is
possible for such light systems. Aiming at the study of the
continuum effects, that increase significantly the computational
load, here we truncate the shell-model space to include only
seniority $s=0$ and $s=1$ states. This method, ``exact pairing +
monopole", is known \cite{EP} to work well for shell model systems
involving only one type of nucleons (in the case of the oxygen
isotope chain only neutrons and the interaction matrix elements
with isospin $T=1$ are involved). The two important ingredients of
residual nuclear forces are treated  by this method exactly: the
monopole interaction that governs the binding energy behavior
throughout the mass region, and pairing that is responsible for
the emergence of the pair condensate, renormalization of
single-particle properties and collective pair vibrations.

\begin{figure}
\begin{center}
\includegraphics[width=14 cm]{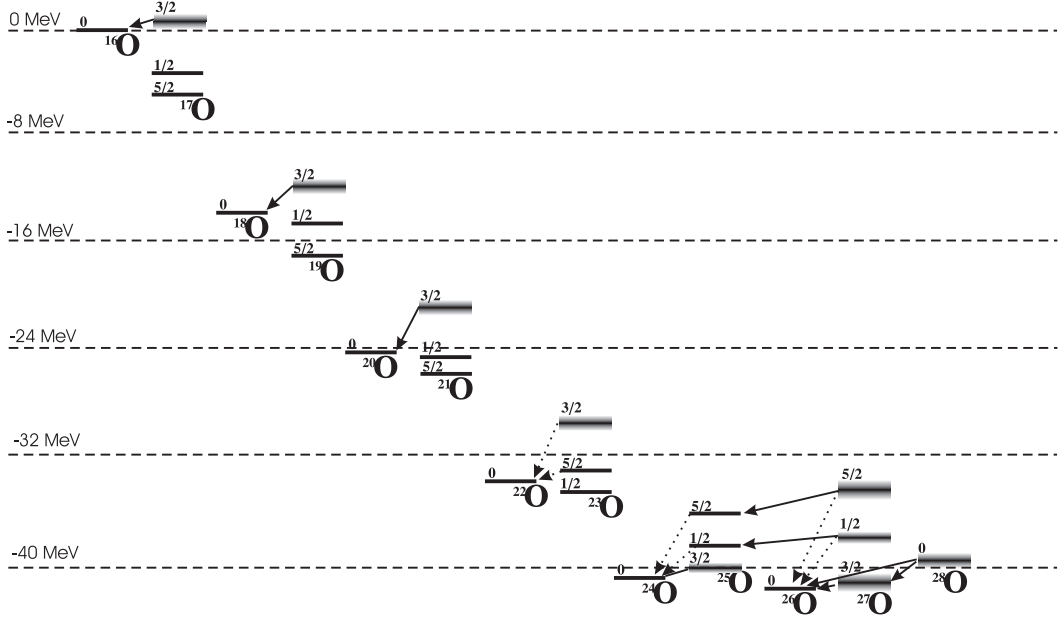}
\end{center}
\caption{\label{oscheme} Schematic picture of lowest seniority
states and possible decays in oxygen isotopes. Dotted arrows
indicate the decays blocked by seniority conservation.}
\end{figure}

In the resulting simplified shell-model description, the set of
the original 30 two-body matrix elements in the $T=1$ channel is
reduced to 12 most important linear combinations. Six of these are
the two-body matrix elements for pair scattering in the $L=0$
channel describing pairing, and the other six correspond to the
monopole force in the particle-hole channel,
\begin{equation}
\overline{V}_{j,j'}\equiv\sum_{L\ne 0} (2L+1) \langle
(j,j')L1|V|(j,j')L1\rangle,                      \label{mono}
\end{equation}
where $j$ and $j'$ refer to one of the three single-particle
levels.

We assume here that the $0d_{3/2}$ orbital belongs to the
single-particle continuum and therefore its energy has an
imaginary part. In this model we account for two possible decay
channels $|c\rangle$ for each initial state $|\Phi\rangle$, a
one-body channel, $c=1$, and a two-body channel, $c=2$. The
one-body decay changes the seniority of the $0d_{3/2}$ orbital
from 1 to 0 in the decay of an odd-$A$ nucleus and from 0 to 1 for
an even-$A$ nucleus. The two-body decay with the zero angular
momentum of the pair removes two paired particles and does not
change the seniority. The two channels lead to the lowest energy
state of allowed seniority in the daughter nucleus; transitions to
excited pair-vibrational states are ignored. This results in
\begin{equation}
e_{3/2}(\Phi)=\epsilon_{3/2}-\frac{i}{2} {\alpha_{3/2} }\,
(E_{\Phi} -E^{(1)})^{5/2}
 -i \,\alpha_{3/2}(E_{\Phi}-E^{(2)})^{5/2}\,,  \label{Ox}
\end{equation}
where we assumed that one- and two-body decay parameters
$\gamma^{(c)}_{j}$ are related as
$\gamma^{(1)}_{3/2}=\gamma^{(2)}_{3/2}/2\equiv\gamma_{3/2}$, and
the particles are emitted in the $d$-wave with $\ell=2\,$. The
energy dependence of the widths near decay thresholds $E^{(c)}$ is
very important; $\alpha_{3/2}$ is the reduced width parameter that
differs from $\gamma_{3/2}$ by the absence of the energy factor.
Three states with the valence particle at one of the
single-particle orbitals can be identified as the $5/2^{+}$ ground
state and $1/2^{+}$ and $3/2^{+}$ excited states in the spectrum
of $^{17}$O. Their energies relative to $^{16}$O correspond to the
single-particle energies in the USD model. Experimental evidence
indicates that the $3/2^{+}$ state decays via neutron emission
with the width $\Gamma(^{17}$O$)=96$ keV. This information allows
us to fix our parameter $\alpha_{3/2}=
\Gamma(^{17}$O$)/(\epsilon_{3/2})^{5/2}=0.028$ (MeV)$^{-3/2}\,.$
Other two states are particle-bound,
$\gamma_{1/2}=\gamma_{5/2}=0\,.$

With complex single-particle energies, the non-Hermitian effective
Hamiltonian for the many-body system is constructed in a regular
way \cite{MW,Sok}. The Hamiltonian includes the Hermitian pairing
and monopole terms as well as the energy-dependent non-Hermitian
effective interaction through the open decay channels, the
structure of which is dictated by unitarity \cite{Sok}. Energy
dependence is determined by the proximity of thresholds, Eq.
(\ref{Ox}). We move along the chain of isotopes starting from
$^{16}$O. In this way, for each $A$ the properties of the possible
daughter systems $A-1$ and $A-2$ are known. The chain of isotopes
under consideration is shown in Fig. {\ref{oscheme}, which
includes $s=0$ and $s=1$ states and indicates possible decays. The
decays indicated by the dotted arrows are blocked in our model due
to the exact seniority conservation. Non-pairing interactions in
the full shell model mix seniorities making these decays possible.
Since the effective Hamiltonian depends on energy and all
threshold energies have to be determined self-consistently, we
solve this extremely non-linear problem iteratively. We start from
the shell-model energies $E_{{\rm s.m.}}$ corresponding to a
non-decaying system. Then the diagonalization of the Hamiltonian
at this energy allows us to determine the next approximation to
the complex energies ${\cal E}=E-(i/2)\Gamma$ that give the
position and the width of a resonance. This cycle is repeated
until convergence that is usually achieved in less than ten
iterations.

The results of the calculations and comparison with experimental
data for oxygen isotopes are shown in Table \ref{tab:oxygen}.
Despite numerous oversimplifications related to seniority
truncation, limitations on the configuration mixing and
restrictions on possible decay channels and final states, the
overall agreement observed in Table \ref{tab:oxygen} is quite
good. If experimental data are not available, the results can be
considered as predictions. The main merit of this calculation is
in demonstrating the power and practicality of the EP method
extended to continuum problems. The same calculation predicts also
\cite{ZV} the cross sections of the processes related to the
included channels providing the unified description of the
structure and reactions with loosely bound nuclei.

\begin{table}
$$
\begin{array}{|r|r|r|r||r|r|}
\hline
A& J & E\, {\rm (MeV) }& \Gamma \, {\rm (keV) } & E_{\rm exp.} \, {\rm (MeV) }& \Gamma_{\rm exp.} \, {\rm (keV) } \\
\hline
16& 0& {\bf 0.00} & 0 &  {\bf 0.00} & 0 \\
17& 5/2 & {\bf -3.94 } & 0 & {\bf -4.14} & 0 \\
17& 1/2& 0.78& 0 & 0.87 & 0 \\
17& 3/2& 5.59& 96 & 5.08 & 96 \\
18& 0 & {\bf -12.17}& 0 & {\bf -12.19} & 0 \\
19& 5/2 & {\bf -15.75} & 0 & {\bf  -16.14} & 0 \\
19& 1/2& 1.33& 0 & 1.47 & 0 \\
19& 3/2& 5.22& 101 & 6.12 & 110 \\
20& 0& {\bf -23.41} & 0 & {\bf -23.75} & 0 \\
21& 5/2 & {\bf -26.67} & 0 & {\bf -27.55} & 0 \\
21& 1/2& 1.38& 0 & & \\
21& 3/2& 4.60& 63 & &  \\
22& 0& {\bf -33.94 }& 0 & {\bf -34.40 } & 0 \\
23& 1/2 & {\bf -35.78} & 0 & {\bf -37.15} & 0 \\
23& 5/2 & 2.12& 0 & &  \\
23& 3/2 & 2.57& 13 & &  \\
24& 0& {\bf -40.54} & 0 &{\bf -40.85} & 0\\
25& 3/2& {\bf -39.82}& 14 & &  \\
25& 1/2& 2.37& 0 &  &  \\
25& 5/2 & 4.98& 0 & &  \\
26& 0& {\bf -42.04} & 0  & & \\
27& 3/2 & {\bf -40.29} & 339  & & \\
27& 1/2& 3.42& 59 &  &  \\
27& 5/2 & 6.45& 223 & &  \\
28& 0& {\bf -41.26} & 121 &  & \\
\hline
\end{array}
$$
\caption{Seniority $s=0$ and $s=1$ states in oxygen isotopes.
Energies and neutron decay widths are shown. Results are compared
to the known data. Ground state energies relative to the $^{16}$O
core are given in bold. The rest of the energies are excitation
energies in a given nucleus. \label{tab:oxygen}}
\end{table}

\section{Thermal properties}

\subsection{BCS approach}

The properties of the dense spectrum of highly excited states are
usually described in statistical terms of level density, entropy
and temperature. The shell model analysis \cite{big,ann} revealed
a certain similarity between many-body quantum chaos and
thermalization. In particular, the Fermi-liquid approach to the
complex many-body system modelled as a gas of interacting
quasiparticles, turns out to be applicable not only in the
vicinity of the Fermi surface but even at high excitation energy.

Here we consider the thermalization properties of the paired
system. The BCS operates with the quasiparticle thermal ensemble.
The expectation value for an occupancy of a given single-particle
state is
\begin{equation}
n=\langle a^{\dagger}a\rangle=u^{2}\langle\alpha^{\dagger}
\alpha\rangle + v^{2}\langle\tilde{\alpha}
\tilde{\alpha^{\dagger}}\rangle + uv\langle\alpha^{\dagger}
\tilde{\alpha^{\dagger}}+\tilde{\alpha}\alpha\rangle. \label{th2}
\end{equation}
Under the assumption of the thermal equilibrium quasiparticle
distribution, the last term in (\ref{th2}) disappears, while the
first and the second terms give
\begin{equation}
\langle\alpha^{\dagger}_{j}\alpha_{j}\rangle = \nu_{j}, \quad
\langle\tilde{\alpha}_{j}\tilde{\alpha_{j}^{\dagger}}\rangle=
1-\nu_{\tilde{j}}.                            \label{equil}
\end{equation}
The occupation numbers for quasiparticles are defined by the Fermi
distribution with the zero chemical potential,
\begin{equation}
\nu_{j}(T)=[1+\exp(\beta e_{j})]^{-1}, \quad \beta=1/T, \label{q}
\end{equation}
and temperature-dependent quasiparticle energies
\begin{equation}
e_{j}(T)=\sqrt{(\epsilon_{j}-\mu)^{2}+\Delta_{j}^{2}(T)},\label{e}
\end{equation}
The thermal evolution of the pairing gap $\Delta(T)$ is determined
by the self-consistent BCS equation with the quasiparticle
blocking factor included,
\begin{equation}
\Delta_{j}(T)=\sum_{j'}G_{jj'}\frac{[1-\nu_{j'}(T)-
\nu_{\tilde{j'}}(T)]\Delta_{j'}(T)}{2e_{j'}(T)}. \label{tgap}
\end{equation}
The occupation numbers of original particles are given by
\begin{equation}
n_{j}=u_{j}^{2}\nu_{j}+v_{j}^{2}(1-\nu_{j}),       \label{tn}
\end{equation}
where the coherence factors $u$ and $v$ also depend on temperature
via $e_{j}(T)$. (This discussion is closely related to Ref.
\cite{dangandvz}.) As a result, we obtain equations for the gap
(\ref{tgap}) and chemical potential using Eq. (\ref{tn}) at a
given external temperature $T$ that governs the quasiparticle
distribution.

\subsection{Statistical spectroscopy of pairing}

A form of the pairing Hamiltonian allows for a relatively simple
calculation of its spectroscopic moments. In this section we limit
our consideration to the zero seniority block of a ladder system
of total capacity $\Omega$ with double-degenerate orbitals; the
generalization for more realistic cases is straightforward. In the
ladder system the diagonal matrix elements simply renormalize
single-particle energies. It is convenient to set the chemical
potential to zero and use variables $\varepsilon_1=\epsilon_1-G_{1
1}/2$ following Eq. (\ref{eprime}). We also denote the
off-diagonal pairing matrix elements as ${\cal G}_{1 2}=
(1-\delta_{1 2}) G_{1 2}$.

The centroid of the distribution is determined by the
single-particle spectrum,
\begin{equation}
\langle \langle E \rangle\rangle = N \overline{\epsilon}\,,
                                            \label{ro1}
\end{equation}

Here the double brackets imply averaging over all many-body
states, while the overline means averaging over single-particle
states according to the definition
\begin{equation}
\overline{{\cal G}^k}=\frac{2}{\Omega}\,{\rm Tr}({\cal G}^k).
                                           \label{over}
\end{equation}
The second moment of the distribution, the variance, is a sum in
quadratures of the single-particle width and the width due to
pairing,
\begin{equation}
\sigma^2=\langle\langle {(E-\langle \langle E \rangle \rangle})^2
\rangle\rangle= \frac{2N(\Omega-N)}{\Omega-2} \, \left(
\overline{(\varepsilon- \overline{\varepsilon})^2} +
\frac{1}{4}\overline{{\cal G}^2} \right ).      \label{ro2}
\end{equation}
The third moment, the skewness, indicates deviations from the
normal distribution. It is given by
\begin{equation}
\langle \langle{(E-\langle\langle E \rangle\rangle })^3
\rangle\rangle =-\frac{N(\Omega-N)(\Omega-N-2)}{(\Omega-2)
(\Omega-4)} \,\overline{{\cal G}^3} \,.              \label{ske}
\end{equation}
All odd central moments are asymmetric in ${\cal G}$ and thus
vanish for ${\cal G}=0.$ The skewness is also a special case since
in the ladder system it does not depend on single-particle
energies. For attractive pairing the skewness is always negative
indicating a longer tail of the distribution towards the lower
energies. This supports the pairing character of the low-lying
states that due to their collective nature are pushed further down
from the centroid of the distribution.

The density of states $\rho(E)$ allows for a thermodynamical
determination of the temperature,
\begin{equation}
\frac{1}{T}=\frac{\partial}{\partial E}\ln \left(\rho(E)\right).
                                           \label{tdef}
\end{equation}
Despite the presence of higher moments, the density of states of
paired systems, as in a more general class of two-body
Hamiltonians \cite{big}, can be closely approximated by the
Gaussian distribution. An actual distribution $\rho (E)$ is shown
in Fig. \ref{ro}. Assuming the Gaussian distribution with the mean
value (\ref{ro1}) and the width $\sigma$, Eq. (\ref{ro2}), the
temperature as a function of energy can be found \cite{big} using
(\ref{tdef}),
\begin{equation}
T(E)=\frac{\sigma^2}{\langle \langle E \rangle\rangle -E}.
                                               \label{temp}
\end{equation}
The negative $T$ branch is an artifact of the finite Hilbert
space.

 The Gaussian distribution is getting distorted when the
pairing becomes strong and the low-lying states very collective. A
minor manifestation of this collectivity is seen in Fig.
\ref{ro}(b) for $G=1$. As $G$ grows, the deviations from the
Gaussian shape become more transparent clearly revealing the
seniority structure as seen in Fig. \ref{rol14G3}.

\begin{figure}
\begin{center}
\includegraphics[width=14.0 cm]{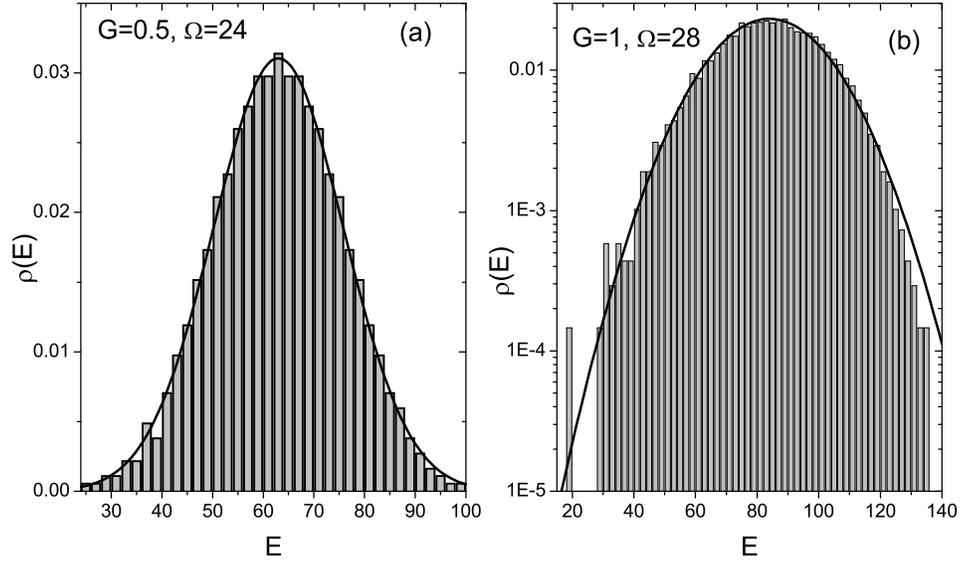}
\end{center}
\caption{\label{ro} Density of states as a function of energy is
shown by the histogram. The solid curve indicates the Gaussian
distribution with the parameters determined by Eqs. (\ref{ro1})
and (\ref{ro2}). Both panels refer to half-occupied systems. Panel
(a) for the 12-level system with $G=0.5$ displays an overall good
Gaussian fit while the deviations are more clear in panel (b)
where the density of states is plotted in the log scale for a
larger system with stronger pairing, $G=1$. Here the effect of
negative skewness is visible, as well as the presence of the gap.
}
\end{figure}
\begin{figure}
\begin{center}
\includegraphics[width=14.0 cm]{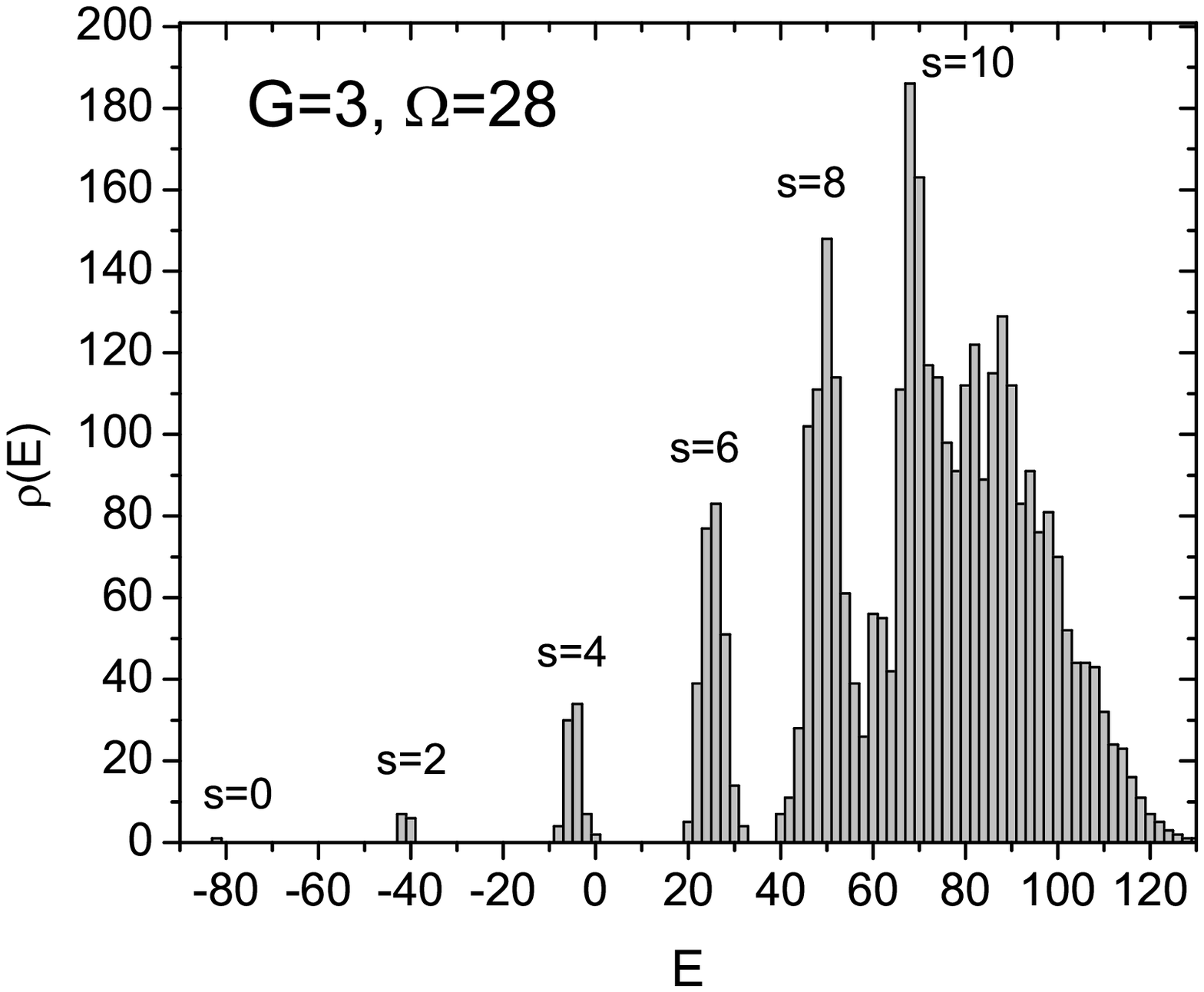}
\end{center}
\caption{\label{rol14G3} Density of states in the half-occupied
system with $\Omega=28$ and pairing strength $G=3.$ The strong
pairing dominates single-particle structure resulting in the peaks
of the level density that correspond to the quasispin families of
the degenerate model. }
\end{figure}

\subsection{Quasiparticle temperature}

As was discussed in detail in \cite{big,ann}, there is no unique
definition of temperature in a self-sustaining isolated mesoscopic
system. Complementary to the thermodynamic (microcanonical)
definition of the previous subsection, we can find the effective
value of temperature for each individual many-body state by
fitting the occupation numbers found in the EP solution to those
given by the thermal ensemble, Eq. (\ref{tn}). We can refer to
this assignment as a measurement with the aid of the quasiparticle
thermometer, and denote the resulting temperature as ${\cal T}\,.$

The correspondence between excitation energy and quasiparticle
temperature ${\cal T}\,$ for each eigenstate in the 12-level model
is presented in Fig. \ref{ETl12}. The scattered points clearly
display a regular trend to thermalization in agreement with the
hyperbolas predicted by Eq. (\ref{temp}). Although thermodynamic
and quasiparticle temperature are well correlated, Fig.
\ref{TTl12}, their numerical scales are different, $T \approx 2.5
{\cal T}.$ Furthermore, the concept of quasiparticle
thermalization is meaningful only for relatively weak pairing,
where large fluctuations due to the proximity of the phase
transition are present. The quality of thermalization deteriorates
as stronger pairing makes the dynamics more and more regular. The
inability of the strongly paired system to fully thermalize
dynamics was demonstrated earlier \cite{vran}. The role of
non-pairing interactions is essential for equilibration. But the
failure of the single-particle thermometer to reflect correctly
the spectral evolution in the limit of very strong interaction is
a general feature \cite{big,ann}.

\begin{figure}
\begin{center}
\includegraphics[width=14 cm]{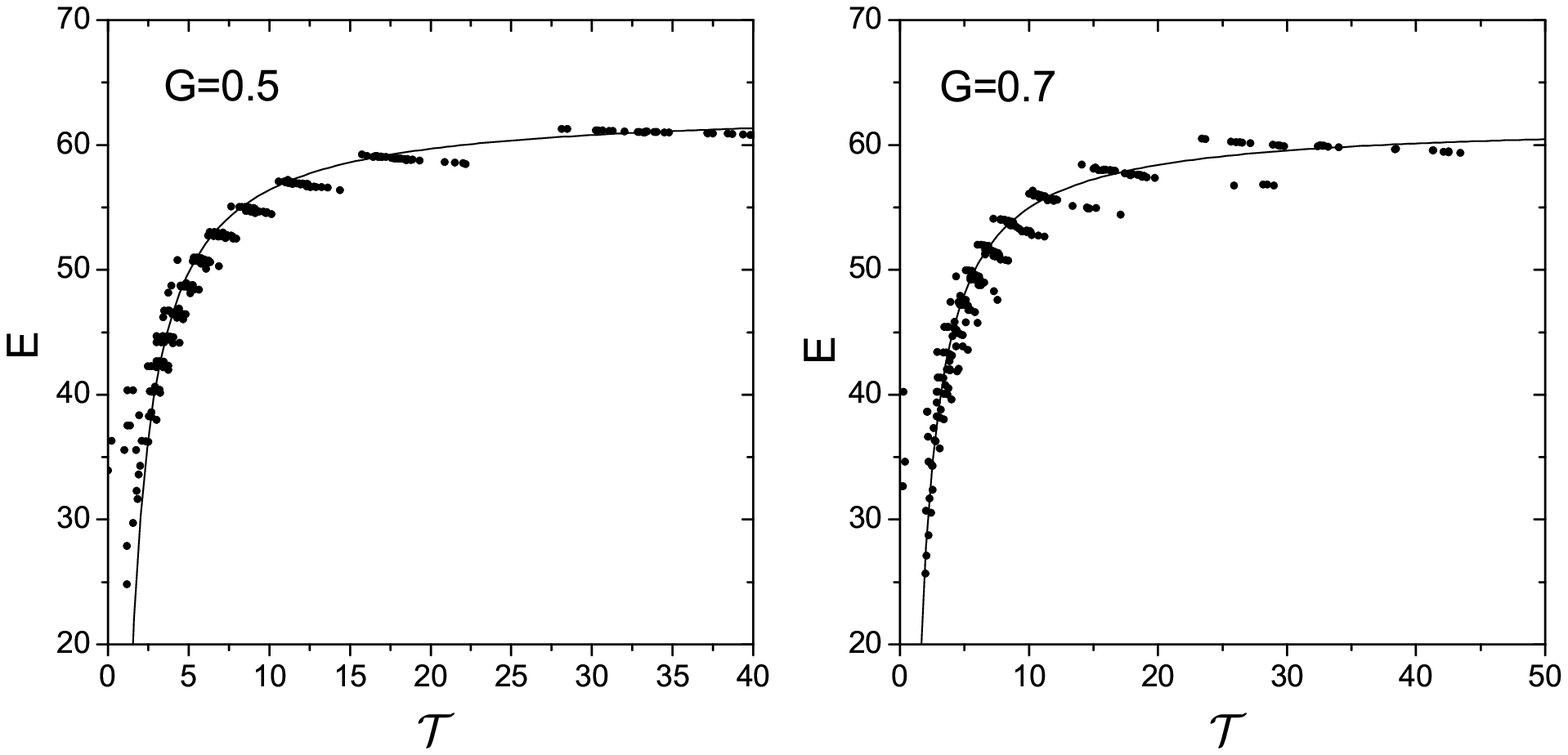}
\end{center}
\caption{\label{ETl12}Excitation energy and quasiparticle
temperature ${\cal T}$ of individual states in the half-occupied,
12-level ladder model. The solid line shows the energy-temperature
relation from Eq. (\ref{temp}), assuming scaling $T \approx 2.5
{\cal T}.$
 }
\end{figure}

\begin{figure}
\begin{center}
\includegraphics[width=14 cm]{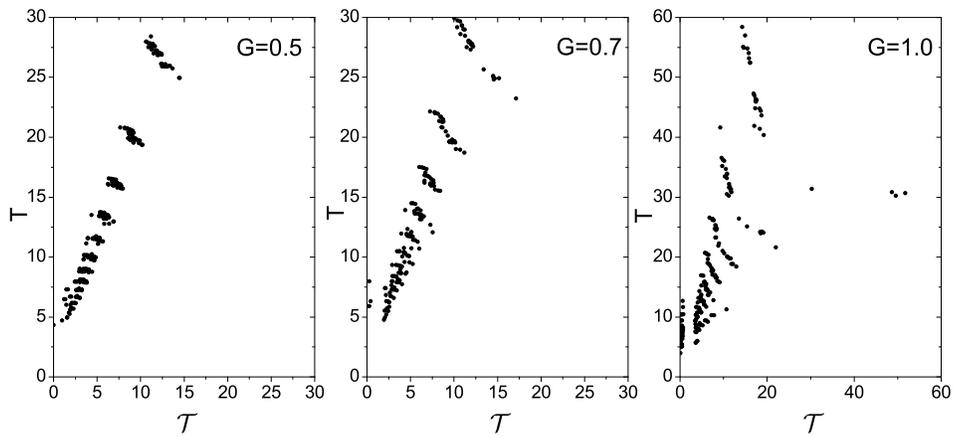}
\end{center}
\caption{\label{TTl12}Comparison of thermodynamic and
quasiparticle temperatures for the half-occupied, 12-level ladder
model. Three panels correspond to pairing strengths $G=$0.5, 0.7
and 1.0, from left to right, respectively. The regular trends in
scattered points confirm approximate thermalization with
temperature ratio $T \approx 2.5 {\cal T}.$ The agreement
deteriorates for stronger pairing.
 }
\end{figure}

\subsection{Pairing phase transition}

In Fig. \ref{deltatl12} the pairing gap is plotted as a function
of quasiparticle temperature. The gap was calculated using Eq.
(38) with the quasiparticle temperature substituting the BCS
external temperature parameter. This makes the consideration
consistent with the occupancies given by Eqs. (36) and (39). The
half-occupied, 12-level ladder model was again used for this
example. The choice of a larger system not only results in the
increased number of $s=0$ states, but, more importantly, reduces
the particle number fluctuations that can disrupt the fitting
procedure, especially in the pairing phase transition region.

Fig. \ref{deltatl12} demonstrates the phase transition from the
paired state at lower temperature (or excitation energy) to a
normal state at a higher temperature. Few low-lying states have a
considerable pairing gap, whereas the gap disappears in the
sufficiently excited states. The invariant entropy \cite{corent}
can be an alternative method for visualizing the phase transition
\cite{vran,cernaj}. This quantity is basis-independent and
reflects the sensitivity of a particular eigenstate to the changes
of a parameter of the many-body Hamiltonian, here the pairing
strength $G$. The peak in the invariant entropy points out the
location of the pairing phase transition as a function of $G$ and
excitation energy $E$ of a particular state. However, it is
important to notice that the phase transition pattern is strongly
influenced \cite{big,ann} by other parts of residual interaction.

\begin{figure}
\begin{center}
\includegraphics[width=14 cm]{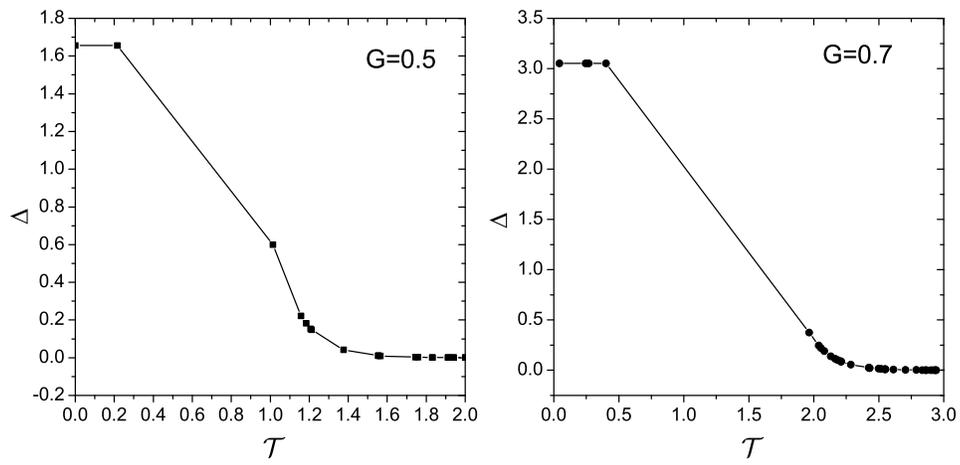}
\end{center}
\caption{Pairing gap as a function of quasiparticle temperature in
the 12-level model with pairing strength $V$=0.5 and 0.7 for left
and right panels, respectively. \label{deltatl12} }
\end{figure}

\section{New theoretical perspectives}

\subsection{Hartree-Fock approximation based on the exact pairing
solution}

Instead of the normal Fermi-occupation picture with a Slater
determinant as a trial many-body function for the mean field
approximation, the EP solution with its specific single-particle
occupancies provides a new starting point for the consistent
consideration of other parts of the residual interaction. In this
subsection we illustrate this point with the help of Belyaev's
\cite{Bel1} pairing plus quadrupole (P+Q) model Hamiltonian for a
single-$j$ level,
\begin{equation}
H=-G \,L^{\dagger} L - \frac{\chi}{2} \sum_{\kappa} {\cal
M}^{\dagger}_{2\,\kappa} {\cal M}_{2\,\kappa}\,, \label{HPandQ}
\end{equation}
where the multipole operators are defined as
\begin{equation}
{\cal M}_{K\,\kappa}=\,\sum_{m_1 m_2}\, (-)^{j-m_1}
\left(\begin{array}{ccc}
      j& K &j\\
      -m_{1}& \kappa &m_{2}\end{array}\right)
\,a^{\dagger}_{2} a_{1}\,.            \label{mult}
\end{equation}
Only the ratio $\chi/G$ of the strength of quadrupole-quadrupole
interaction to the pairing strength is important since the energy
scale can be fixed so that $G=1.$

In the pure pairing limit, $\chi=0$, the degenerate pairing model
is recovered with the ground state energy $(\Omega=2j+1)$
\begin{equation}
E=G \frac{N}{4} (\Omega-N+2).         \label{chi0}
\end{equation}
Pairing correlation energy in the BCS with constant pairing is
given by $\Delta^2/G\,.$ For the exact solution we define the
correlation energy as the ground state expectation value of the
pairing part of the Hamiltonian with the monopole contribution
subtracted. In the degenerate case
\begin{equation}
E_{\rm corr}=G \frac{N}{4} (\Omega-N).   \label{cor}
\end{equation}

The opposite limit with no pairing can be treated by making a
transition to a deformed mean field in the Hartree approximation
\cite{baranger65}. For axially symmetric deformation, the
expectation value of the quadrupole moment is
\begin{equation}
\langle {\cal M}_{2\,0}\rangle=\sum_{m} \frac{2 [3m^2-j(j+1)]}
{\sqrt{\Omega(\Omega^2-1)(\Omega^2-4)}} \, n_m \,, \label{Har}
\end{equation}
in terms of the occupation numbers $ n_m=\langle
a^\dagger_m\,a_m\,\rangle $ in the intrinsic frame with the
$z$-axis oriented along the symmetry axis. In this case
\begin{equation}
\langle {\cal M}_{2\,-2}\rangle=\langle {\cal M}_{2\,2}\rangle=0.
                                                \label{M2}
\end{equation}
The deformed single-particle energies in the body-fixed frame  can
be obtained via the usual self-consistency requirement,
\begin{equation}
\epsilon_m=-\chi\,\frac{2[3m^2-j(j+1)]} {\sqrt{\Omega(\Omega^2-1)
(\Omega^2-2)}}\,\langle{\cal M}_{2\,0}\rangle\,. \label{HFSP}
\end{equation}
The energy minimum for an even-$N$ system corresponds to the Fermi
occupation of the $N/2$ lowest pairwise degenerate orbitals
$|m|=1/2,3/2,\dots (N-1)/2$ for prolate, or $|m|=j,j-1,\dots
j-(N-2)/2$  for oblate shape. The corresponding quadrupole moment
is  then  given by
\begin{equation}
\langle {\cal M}_{2\,0}\rangle
=-\frac{1}{2}\,\frac{N(\Omega^2-N^2)}
{\sqrt{\Omega(\Omega^2-1)(\Omega^2-4)}}               \label{pr}
\end{equation}
for prolate deformation and
\begin{equation}
\langle {\cal M}_{2\,0}\rangle
=\frac{1}{2}\,\frac{N(2\Omega-N)(\Omega-N)}
{\sqrt{\Omega(\Omega^2-1)(\Omega^2-4)}}              \label{ob}
\end{equation}
for oblate deformation. Deformation energy is quadratic in
$\langle {\cal M}_{2\,0}\rangle $, so that the oblate deformation
is preferred for $N<\Omega/2$ and the prolate one for
$N>\Omega/2.$

The full problem is driven by the competition between pairing and
deformation. While deformation tends to split the single particle
energies (Nilsson orbitals), Eq. (\ref{HFSP}), the pairing can
resist such a shape transition by creating a particle distribution
unfavorable for deformation. In the single-$j$ model these effects
have been discussed by Baranger \cite{baranger65} with the help of
the BCS and the Hartree approximation. However, as demonstrated
earlier, the BCS may be unreliable in the transitional region. The
use of the projected HFB and the resulting improvement against
traditional HFB for a similar single-$j$ model have been recently
discussed in \cite{sheikh02}. Our goal here is to supplement the
Hartree treatment with the exact pairing solution.

Similar to the Hartree+BCS approach in \cite{baranger65} we look
for a self-consistent solution, where the occupation numbers in
the deformed basis agree with the exact solution to the pairing
problem \cite{vbz}. Prior to calculations, one can estimate the
ratio $\chi/G$ corresponding to the BCS phase transition. Assuming
for example an oblate shape with $N\le\Omega/2$ we have the Fermi
energy at $m=j-(N-2)/2$ with the density of single-particle states
approximately found as $\nu_{{\rm F}}=2/(\epsilon_{m-1}-
\epsilon_{m})$, that for a half-occupied system leads to
\begin{equation}
(\chi/G )_{\rm crit}\approx 3.6\, \Omega.         \label{crgxi}
\end{equation}

In Fig. \ref{ojhep} we present the results for a model with
$j=19/2$, The particle number, $N=8$, was selected to avoid an
exactly half-occupied shell when particle-hole symmetry and oblate
to prolate shape change lead to special features that are of no
interest for our goal. For this model the transition from
spherical to deformed shape takes place at $\chi/G\approx 70$, Eq.
(\ref{crgxi}). The pairing correlation energy shown in Fig.
\ref{ojhep}(a) as a function of $\chi/G$ starts near $\chi/G=0$
with a value prescribed by the degenerate model. As the relative
strength of the quadrupole-quadrupole interaction increases, the
deformation inhibits pairing. However, in Fig. \ref{ojhep}(a) we
see a key difference between the BCS and exact treatment. Within
the BCS the pairing correlation energy goes to zero quite sharply
once the system becomes deformed. In contrast, the exact solution
finds that pairing correlations decay very slowly and extend far
into the deformed region. In panel \ref{ojhep}(b) the expectation
value of $\langle {\cal M}_{2 0} \rangle$ is shown as a function
of $\chi/G$. In the pairing limit $\chi/G\rightarrow 0$ the
deformation is zero whereas in the deformed limit the value
$\langle {\cal M}_{2 0} \rangle$ is expressed via Eq. (\ref{ob})
or (\ref{pr}). Here again the exact treatment produces a softened
and extended phase transition.

\begin{figure}
\begin{center}
\includegraphics[width=14 cm]{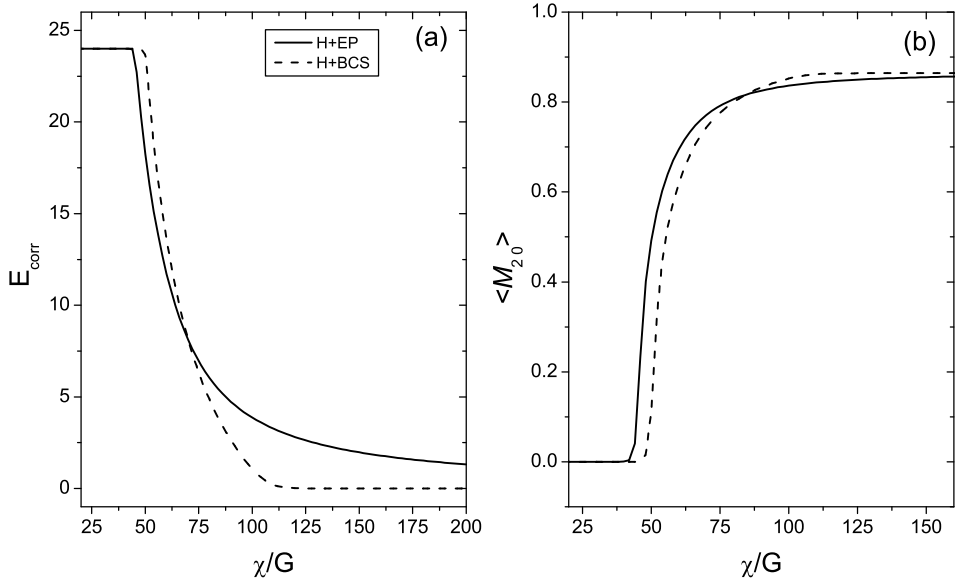}
\end{center}
\caption{Comparison of pairing correlation energy, panel (a), and
quadrupole deformation, panel (b), for Hartree+BCS and Hartree+EP
treatments of the P+Q model for a single-$j$ level, $j=19/2$, and
$N=8$ particles. For pairing correlation energy $G=1.0\,$ is
assumed. \label{ojhep} }
\end{figure}

The Hartree approximation ignores another important effect
relevant to our consideration, namely, the contribution of the
exchange terms to the pairing channel. This contribution is
particularly strong in small systems and leads to an additional
enhancement of the pairing strength $G\rightarrow G+2\chi/\Omega^2$
\cite{volya02}. The results of Hartree-Fock + EP calculations that
include this additional term are shown in Fig. \ref{ojhfep}. Due
to the exchange term, pairing correlations never disappear.
Presence of pairing correlations in a pure quadrupole-quadrupole
Hamiltonian is also confirmed by an exact solution in the full
shell model diagonalization \cite{volya02}. The BCS treatment,
however, fails to reproduce this effect.

\begin{figure}
\begin{center}
\includegraphics[width=14 cm]{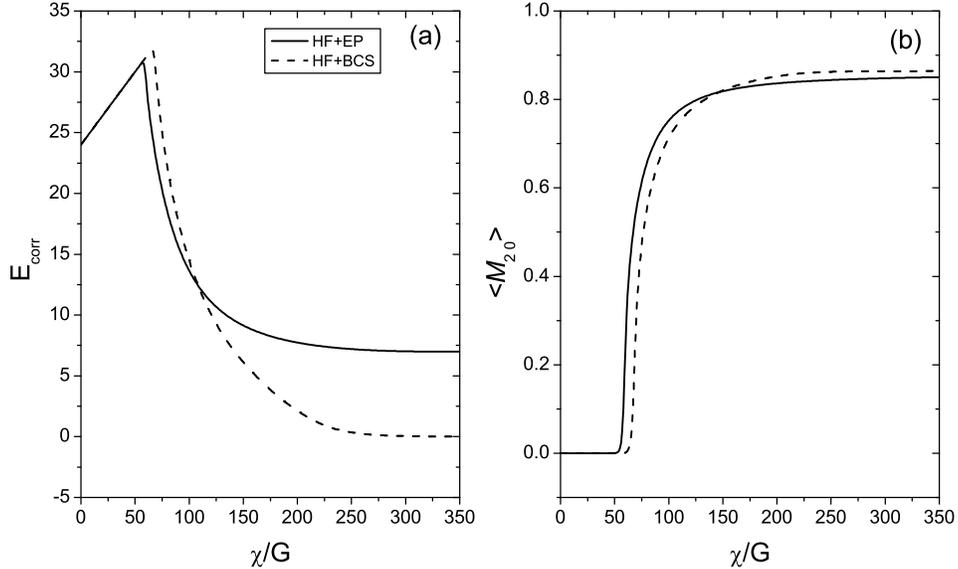}
\end{center}
\caption{\label{ojhfep} Comparison of pairing correlation energy
and quadrupole deformation; in contrast to Fig.~\ref{ojhep}, the
exchange contribution from the quadrupole-quadrupole interaction
to the pairing channel is included. }
\end{figure}

\subsection{New Random Phase Approximation}

Here we show how the RPA-like approximation can be developed
starting from the exact solution of the pairing problem. There are
two main types of the RPA used in the literature (and a variety of
close approaches distinguished by the details of the formalism),
the RPA based on the vacuum of noninteracting particles or that of
the BCS, or HFB, quasiparticles (the so-called QRPA). We will try
to describe collective vibrations generated by the residual
interaction on top of the exact ground state of the pairing
problem. The formalism of the generalized density matrix
\cite{GDM,Jap} seems to be suitable for this purpose.

The generalized density matrix (GDM) is the set of the operators
\begin{equation}
R_{12}=a^{\dagger}_{2}a_{1}                     \label{R1}
\end{equation}
acting in the full Hilbert space of a many-body system; this set
at the same time forms a matrix labeled by single-particle
subscripts ({\sl 1,2}). We do not perform any canonical
transformation to the quasiparticle operators and therefore work
invariably within a system of a certain particle number. The
one-body observables as operators in many-body space are traces of
the GDM operator over single-particle indices,
\begin{equation}
Q=\sum_{a}q_{a}\quad\Rightarrow\quad
Q=\sum_{12}q_{12}a^{\dagger}_{1}a_{2}={\rm Tr}\,(qR). \label{R2}
\end{equation}
With the Hamiltonian of the system taken as a sum of independent
particle energies in the mean field, $\epsilon_{1}$, and the
general residual two-body interaction $V_{12;34}$, the exact
operator equations of motion for the GDM can be symbolically
written as
\begin{equation}
[R,H]=[S,R],                                      \label{R3}
\end{equation}
where $S$ is the generalized self-consistent field operator (a
linear functional of the GDM),
\begin{equation}
S=\epsilon + W\{R\}, \quad W_{14}\{R\}=\sum_{23}V_{12;34}R_{32},
                                                   \label{R4}
\end{equation}
and the interaction matrix elements are antisymmetrized.

Now we assume that the Hamiltonian contains the pairing part
(\ref{BCS:ham}) as well as other residual interactions,
$H=H_{p}+H'$. Correspondingly we can put
\begin{equation}
R=R^{\circ}+R', \quad W=W^{\circ}+W', \quad
W^{\circ}=W\{R^{\circ}\}, \quad W'=W\{R'\}.     \label{R5}
\end{equation}
The assumption of the exact solution of the pairing problem means
that we found the occupancies, $R^{\circ}$, and the pairing
potential, $W^{\circ}$, satisfying
\begin{equation}
[R^{\circ},H_{p}]=[\epsilon+W^{\circ},R^{\circ}].   \label{R6}
\end{equation}
This stage of the solution provides the states $|s,a\rangle$ with
seniority $s$ and energy $E_{sa}$, where $a$ numbers the states
within the subset of certain seniority; if needed we also can
explicitly indicate rotational quantum numbers $J_{a}$ and
$M_{a}$. The remaining part $R'$ of the GDM should satisfy
\begin{equation}
[R',H_{p}]+[R^{\circ},H']+[R',H']=[W',R^{\circ}]+[\epsilon+
W^{\circ},R']+[W',R'].                           \label{R7}
\end{equation}
The commutators in such expressions are to be understood as, for
example
\begin{equation}
[W',R']_{12}=\sum_{3}(W'_{13}R'_{32}-R'_{13}W'_{32}). \label{R8}
\end{equation}
This is the point where we can make RPA-like approximations.

For definiteness, we consider the transitions from the paired
states $s=0,J=0$ to the states with $s=2,J\neq 0$ in the next
sector. We are looking for a collective mode that is related to
such excitations. This means that there exist states, in our case
coherent combinations of excited states with certain $J$, that
have large off-diagonal matrix elements of excitation by a
one-body multipole operator from the ground state. The latter can
be in turn renormalized by the collective mode. Let us
characterize this branch of the spectrum with the help of
collective coordinates $\alpha$ and conjugate momenta $\pi$ (we
omit in this symbolic derivation their quantum numbers of angular
momentum and its projection). These variables are Hermitian
quantum operators that satisfy the commutation relation
$[\alpha,\pi]=i$ so that no procedure of subsequent requantization
is needed. The collective Hamiltonian of the mode can be written
as
\begin{equation}
H'=\frac{1}{2}C\alpha^{2}+ \frac{1}{2B}\pi^{2}+\,\dots, \label{R9}
\end{equation}
where the scalar contraction of the tensor operators is implied
and the dots include high order anharmonic terms important for the
soft mode \cite{jol,wyom}.

We are looking for the operator solution of Eq. (\ref{R7}) in the
form of the expansion in collective operators $\alpha$ and $\pi$,
\begin{equation}
R'_{12}=r^{(10)}_{12}\alpha+r^{(01)}_{12}\pi+\,\dots, \label{R10}
\end{equation}
and
\begin{equation}
W'_{12}=w^{(10)}_{12}\alpha+w^{(01)}_{12}\pi+\, \dots, \label{R11}
\end{equation}
where the superscripts $(n,m)$ refer to the component containing
$n$ collective coordinate and $m$ collective momenta operators.
The dots again denote the higher order parts, $n+m>1$, symmetrized
in due way \cite{jol,wyom}. The collective operators producing the
transition in many-body space are written explicitly whereas the
coefficients $r^{(nm)}_{12}$ and $w^{(nm)}_{12}$ are the
$c$-numbers to be found as single-particle amplitudes of the
coherent superposition that forms a collective mode. The operator
expansion (\ref{R10}, \ref{R11}) does not assume the smallness of
anharmonic effects - we merely decompose the problem in various
operator structures. In the present context we limit ourselves by
the harmonic part although it can be used \cite{jol,wyom} for the
situations of strong anharmonicity as well.

The operator $R$, by definition (\ref{R1}), has seniority
selection rules $|\Delta s|=2$. We take in the operator equation
(\ref{R7}) matrix elements $\langle 0|c\rangle$ between the ground
state $|0\rangle$ and the collective state $|c\rangle$ in the
adjacent sector $s=2$ with angular momentum corresponding to that
of collective operators $\alpha$ and $\pi$. Now we evaluate the
matrix elements of various terms in Eq. (\ref{R7}) aiming at the
segregation of terms linear in $\alpha$ and $\pi$. The first term
in the right hand side gives, according to Eq. (\ref{R10}),
\begin{equation}
[R'_{12},H_{p}]=(\bar{E}^{\circ}_{c}-
\bar{E}^{\circ}_{0})(r^{(10)}_{12}\alpha+ r^{(01)}_{12} \pi).
                                                   \label{R12}
\end{equation}
Here the barred energies are the centroids of the energy
distribution of the actual ground state and the one-phonon state
in the sectors $s=0$ and $s=2$, respectively. The second term in
eq. (\ref{R7}) does not have required matrix elements whereas in
the commutator $[R',H']$ we need to perform commutation explicitly
using the assumed form of the collective operators
(\ref{R10}-\ref{R12}),
\begin{equation}
[R'_{12},H']=\frac{i}{B}r^{(10)}_{12}\pi -iCr^{(01)}_{12}\alpha.
                                                 \label{R13}
\end{equation}

The situation with the terms $[W^{\circ},R']$ and $[W',R^{\circ}]$
is more complicated. Within each sector of given $s$, the pairing
solution GDM $R^{\circ}$ has not only diagonal but also
off-diagonal elements between the eigenstates of $H_{p}$. As shown
in Figs. 2 and 3, very few pair-vibrational states have
significant off-diagonal matrix elements of this type. For our
illustrative purposes, here we neglect the off-diagonal terms
within a given sector and take into account only diagonal elements
of $R^{\circ}$ and $W^{\circ}$. The neglected contributions
correspond to the anharmonic admixtures of pair vibrations to
multipole modes and can be easily included in the consideration.
Because of the specific character of the monopole pairing
interaction, the matrix elements of $R^{\circ}$ and $W^{\circ}$
are diagonal over single-particle subscripts as well. Higher order
structures in the collective Hamiltonian and in the GDM, as well
as terms generated by the commutator $[W',R']$ do not contribute
to matrix elements linear in $\alpha$ and $\pi$ and with the
selection rule $\Delta s=2$. But, similarly to Eq. (\ref{R12}),
the commutators with $R^{\circ}$ and $W^{\circ}$ bring in the
differences of the single-particle occupancies and pairing
potentials averaged over the states contributing to the collective
mode.

As a result, we come to the coupled equations for the coordinate
and momentum RPA amplitudes (these contributions can be
distinguished by their behavior under time-reversal operation in
the sector with $J\neq 0$):
\begin{equation}
r^{(10)}_{12}\Omega_{12}-iCr^{01}_{12}=
[\bar{n}_{2}(c)-\bar{n}_{1}(0)]w^{10}_{12},       \label{R14}
\end{equation}
\begin{equation}
r^{01}_{12}\Omega_{12}+\frac{i}{B}r^{10}_{12}=
[\bar{n}_{2}(c)-\bar{n}_{1}(0)]w^{01}_{12}.      \label{R15}
\end{equation}
Here the generalized frequencies are introduced,
\begin{equation}
\Omega_{12}=\bar{E}^{\circ}_{0}-\bar{E}^{\circ}_{0}+
\epsilon_{2}-\epsilon_{1}+\bar{W}^{\circ}_{c}-\bar{W}^{\circ}_{0}.
                                                 \label{R16}
\end{equation}
This set of equations leads to a formal solution, analogous to
that in the conventional RPA,
\begin{equation}
r^{10}_{12}=\frac{[\bar{n}_{2}(c)-\bar{n}_{1}(0)]}{\Omega^{2}_{12}-
\omega^{2}}(\Omega_{12}w^{10}_{12}+iCw^{01}_{12}), \label{R17}
\end{equation}
\begin{equation}
r^{01}_{12}=\frac{[\bar{n}_{2}(c)-\bar{n}_{1}(0)]}{\Omega^{2}_{12}-
\omega^{2}}(\Omega_{12}w^{01}_{12}+\frac{i}{B}w^{10}_{12}),
                                              \label{R18}
\end{equation}
where the unknown collective frequency is $\omega=(C/B)^{1/2}$.

The collective elements of the GDM are to be found from the
integral equations (\ref{R17}) and (\ref{R18}) with the specific
choice of the residual interaction self-consistently generating
the field $W'$, Eq. (\ref{R5}). This can be done explicitly in the
case of the factorizable multipole-multipole force; the frequency
$\omega$ and correct normalization of the mode are also obtained
in this process similarly to the standard procedure in terms of
the barred quantities. Having at our disposal the phonon
amplitudes $r^{(10)}_{12}$ and $r^{(01)}_{12}$, we can
self-consistently find the barred quantities averaged over the
collective wave functions. This procedure, that reminds the
thermal RPA built on the equilibrium density matrix but with
occupation numbers and mean-field corrections defined by the
interaction rather than by an external heat bath, can be performed
in an iterative manner. The cranking description for the deformed
nucleus can be also reformulated in the same spirit; it is
interesting to note that the pure EP solution predicts \cite{vran}
an yrast-line with the moment of inertia close to the rigid body
value.

\section{Conclusion}

The pioneering work by Belyaev \cite{Bel} carried out a detailed
analysis of pairing phenomena in nuclei. Applying the BCS
techniques in the nuclear shell model environment, he demonstrated
the effects of pairing on various nuclear properties, including
the ground state structure, single-particle transitions,
collective vibrations, onset of deformations and rotations. It was
shown that the Cooper phenomenon in systems with a discrete
single-particle spectrum does require, in contrast to large
systems, a certain strength of the pairing interaction. The
drawbacks of the BCS approximation, related to the particle number
violation and the sharp disappearance of pairing correlations at
the phase transition point, were also pointed out.

The development started with Belyaev's work and supported by
similar studies \cite{Sol,Migd} was continued through the next
forty years. Now the pairing problem is well and alive being one
of the main chapters of modern nuclear physics and mesoscopic
physics in general. The interest in pairing is constantly revived
by the accumulation of data and especially by the advances towards
nuclei from stability where the pairing is a key tool that
determines the binding of a system and its response to the
excitation. At this point it becomes increasingly important to get
rid of the shortcomings of the BCS approximation and unify the
description of the structure and reactions.

We presented a way of solving the pairing problem essentially
along the lines similar to that of Belyaev's paper, substituting
the BCS approximation by the exact solution simplified by the
seniority symmetry. As a magnifying glass, this solution reveals
and fixes the weak points of the standard approach. We saw the
importance of the exact treatment for the ground state, low-lying
excitations, coupling through the continuum, and spectroscopic
factors associated with single-particle removal and pair emission
(transfer) reactions. We could also discuss on the new basis the
global properties of the spectrum, thermalization and the phase
transition region. In many cases, this exact treatment of pairing
is practically simpler than solving the BCS equations with
necessary corrections.

Certainly, the pairing problem is only a part of the physics of
strongly interacting self-sustaining systems. Other interactions,
with their own coherent and chaotic features, should be included
in the consideration. We gave preliminary answers to the questions
of further approximations necessary for the cases when the full
problem does not allow for a complete solution. New
generalizations of the mean field approach and random phase
approximations can be developed on the background of the exactly
found paired state. The interplay of pairing and other residual
interactions can be an exciting and practically important topic of
future studies.\\
\\
{\small The inspiring influence of Belyaev's ideas is gratefully
acknowledged by the authors who belong to the first and third
generation of his pupils. The work was supported by the NSF grant
PHY-0070911 and by the U. S. Department of Energy,
Nuclear Physics Division, under contract No. W-31-109-ENG-38.}


\begin{thebibliography}{99}
\bibitem{BM} A. Bohr and B.R. Mottelson, {\sl Nuclear
Structure}, vol. 1 (Benjamin, New York, 1969).
\bibitem{JM} M.G. Mayer and J.H.D. Jensen, {\sl Elementary Theory of
Nuclear Shell Structure} (Wiley, New York, 1955).
\bibitem{Rac} G. Racah, Phys. Rev. {\bf 63}, 367 (1943).
\bibitem{RT} G. Racah, I. Talmi, Phys. Rev. {\bf 89}, 913 (1953).
\bibitem{Judd} B.R. Judd, {\sl Operator Techniques in Atomic Spectroscopy},
(Princeton University Press, 1998).
\bibitem{BCS} J. Bardeen, L.N. Cooper and J.R. Schrieffer, Phys. Rev.
{\bf 106}, 162; {\bf 108}, 1175 (1957).
\bibitem{BMP} A. Bohr, B.R. Mottelson and D. Pines, Phys. Rev. {\bf 110}, 936
(1958).
\bibitem{Bel} S.T. Belyaev, Mat. Fys. Medd. Dan. Vid. Selsk. {\bf 31}, No. 11
(1959).
\bibitem{HFB} A.L. Goodman, Adv. Nucl. Phys. {\bf 11}, 263 (1979).
\bibitem{Bog} N.N. Bogoliubov, JETP {\bf 34}, 58 (1958); Physica {\bf 26}, 1
(1960).
\bibitem{Lipkin60} H.J. Lipkin, Ann. Phys. {\bf 31}, 528 (1960).
\bibitem{Nogami64} Y. Nogami, Phys. Rev. B {\bf 134}, 313 (1964); Y. Nogami and
I.J. Zucker, Nucl. Phys. {\bf 60}, 203 (1964).
\bibitem{Satula} W. Satula and R. Wyss, Nucl. Phys. {\bf A676}, 120 (2000).
\bibitem{sheikh00} J.A. Sheikh and P. Ring, Nucl. Phys. {\bf A665}, 71 (2000).
\bibitem{Klein} G. Do Dang and A. Klein, Phys. Rev. {\bf 143}, 735 (1966).
\bibitem{Bang} J. Bang and J. Krumlinde, Nucl. Phys. {\bf A141}, 18 (1970).
\bibitem{cov} F. Andreozzi, A. Covello, A. Gargano, L.J. Ye, A. and Porrino,
Phys. Rev. C {\bf 21}, 1094 (1980); Phys. Rev. C {\bf 32}, 293
(1985).
\bibitem{Bertsch} K. Hagino and G.F. Bertsch, Nucl. Phys. {\bf A679}, 163
(2000).
\bibitem{big} V. Zelevinsky, B. A. Brown, N. Frazier, and M. Horoi, Phys. Rep.
{\bf 276}, 85 (1996).
\bibitem{Brog} N. Giovanardi, F. Barranco, R.A. Broglia, and E. Vigezzi, Phys.
Rev. C {\bf 65}, 041304 (2002).
\bibitem{Benn} K. Bennaceur, J. Dobaczewski, and M. Ploszajczak, Phys. Rev.
C {\bf 60}, 034308 (1999).
\bibitem{Poves} A. Poves and A.P. Zuker, Phys. Rep. {\bf 70}, 235 (1981).
\bibitem{BW} B.A. Brown and B.H. Wildenthal, Ann. Rev. Nucl. Part. Sci.
{\bf 38}, 29 (1988).
\bibitem{richter91} W. A. Richter, M. G. van der Merwe, R. E. Julies
and B. A. Brown, Nucl. Phys. {\bf A523}, 325 (1991).
\bibitem{wildenthal84}
B. Wildenthal, Prog. Part. Nucl. Phys. {\bf 11}, 5 (1984).
\bibitem{SMMC} Y. Alhassid, D.J. Dean, S.E. Koonin, G. Lang, and W.E. Ormand,
Phys. Rev. Lett. {\bf 72}, 613 (1994).
\bibitem{QMCD} M. Honma, T. Mizusaki, and T. Otsuka, Phys. Rev. Lett. {\bf 77},
3315 (1996).
\bibitem{ECM} M. Horoi, A. Volya, and V. Zelevinsky, Phys. Rev. Lett. {\bf 82},
2064 (1999).
\bibitem{EP} A. Volya, B.A. Brown and V. Zelevinsky, Phys. Lett. B {\bf 509},
37 (2001).
\bibitem{rowe95} H. Chen, T. Song and D.J. Rowe, Nucl. Phys. {\bf A582}, 181
(1995).
\bibitem{yi91} Yang Yi and Xu Gong-ou, Phys. Rev. C {\bf 43}, 1225 (1991).
\bibitem{rowe} D. J. Rowe, {\it Nuclear Collective Motion, Models and Theory}
(Methuen and Co. Ltd., London, 1970).
\bibitem{broglia00} R. A. Broglia, J. Terasaki, and N. Giovanardi, Phys. Rep.
{\bf 335}, 1 (2000).
\bibitem{RingSchuck} P. Ring and P. Schuck, {\sl The Nuclear Many-Body
Problem} (Springer-Verlag, New York, 1980).
\bibitem{dang66} G. Do Dang and A. Klein, Phys. Rev. {\bf 143}, 735 (1966);
{\bf 147}, 689 (1966).
\bibitem{dang68} G. Do Dang, R. M. Dreizler, A. Klein and C. Wu,
Phys. Rev. {\bf 172}, 1022 (1968).
\bibitem{PC} A. Volya and V. Zelevinsky, Preprint MSUCL-1144 (1999).
\bibitem{hogaasen61} J. H\"ogaasen-Feldman, Nucl. Phys. {\bf 28}, 258 (1961).
\bibitem{johns70} O. Johns, Nucl. Phys. {\bf A154}, 65 (1970).
\bibitem{hagino00} K. Hagino and G.F. Bertsch, Nucl. Phys. {\bf A679},
163 (2000).
\bibitem{rowe68} D. J. Rowe, Rev. Mod. Phys. {\bf 40}, 153 (1968).
\bibitem{dukelsky90} J. Dukelsky and  P. Schuck, Nucl. Phys. {\bf A512}, 466
(1990); Phys. Lett. B {\bf 387}, 233 (1996).
\bibitem{passos98}  F. Krmpotic, E.J.V. de Passos, D.S. Delion, J. Dukelsky,
and P. Schuck, Nucl. Phys. {\bf A637}, 295 (1998).
\bibitem{Rich} R.W. Richardson,
Phys. Lett. {\bf 3}, 277 (1963);
Phys. Lett. {\bf 5}, 82  (1963);
Phys. Lett. {\bf 14}, 325 (1965);
J.\ Math.\ Phys.\ {\bf 6}, 1034  (1965);
J.\ Math.\ Phys.\ {\bf 18}, 1802 (1977).
Phys.\ Rev.\ {\bf 141},   949  (1966);
Phys.\ Rev.\ {\bf 144},   874  (1966);
Phys.\ Rev.\ {\bf 159}, 792 (1967).
\bibitem{richardson64} R. W. Richardson and N. Sherman,
Nucl.\ Phys.\ {\bf 52},   221  (1964).
\bibitem{Pitt} J. Dukelsky, C. Esebbag, and S. Pittel, Phys. Rev. Lett.
{\bf 88}, 062501 (2002).
\bibitem{Dyson} F.J. Dyson, J. Math. Phys. {\bf 3}, 166 (1962).
\bibitem{pan98} Feng Pan, J. P. Draayer, and W. E. Ormand, Phys. Lett. B {\bf 422},
1 (1998).
\bibitem{pan00} Feng Pan, J. P. Draayer, and Lu Guo, J. Phys. A {\bf 33},
1597 (2000).
\bibitem{dukelsky00} J. Dukelsky and  P. Schuck, LANL preprint
cond-mat/0009057.
\bibitem{burglin96} O. Burglin and N. Rowley, Nucl. Phys. {\bf A602}, 21
(1996).
\bibitem{molique97} H. Molique and J. Dudek, Phys. Rev. C {\bf 56}, 1795
(1997).
\bibitem{Kerman} A.K. Kerman, R.D. Lawson, and M.H. Macfalane, Phys. Rev.
{\bf 124}, 162 (1961).
\bibitem{Auer} N. Auerbach, Nucl. Phys. {\bf 76}, 321 (1966).
\bibitem{holt98} A. Holt, T. Engeland, M. Hjorth-Jensen, and E. Osnes,
Nucl. Phys. {\bf A634}, 41 (1998).
\bibitem{vran} A. Volya, V. Zelevinsky, and B. A. Brown, Phys. Rev.
C {\bf 65}, 054312 (2002).
\bibitem{fraz} N. Frazier, B.A. Brown, and V. Zelevinsky, Phys.
Rev. C {\bf 54}, 1665 (1996).
\bibitem{gur} V. Zelevinsky, Yad. Fiz. {\bf 65}, 1220 (2002).
\bibitem{hor} M. Horoi, B.A. Brown, and V. Zelevinsky, Phys. Rev.
C {\bf 65}, 027303 (2002).
\bibitem{cov1} V. Zelevinsky and A. Volya, in {\sl Challenges of
Nuclear Structure}, ed. A. Covello (World Scientific, Singapore,
2002) p. 261.
\bibitem{print} M. Horoi, B.A. Brown, and V. Zelevinsky, Phys.
Rev. C, {\sl in press}.
\bibitem{MW} C. Mahaux and H.A. Weidenm\"{u}ller, {\sl Shell-Model
Approach to Nuclear Reactions} (North Holland, Amsterdam, 1969).
\bibitem{Sok} V.V. Sokolov and V.G. Zelevinsky, Nucl. Phys.
{\bf A504}, 562 (1989).
\bibitem{ZV} A. Volya and V. Zelevinsky,
nucl-th/0211039 (2002).
\bibitem{ann} V. Zelevinsky, Ann. Rev. Nucl. Part. Sci. {\bf 46},
237 (1996).
\bibitem{dangandvz} N. Dinh Dang and V. Zelevinsky, Phys. Rev. C
{\bf 64}, 064319 (2001).
\bibitem{corent} V.V. Sokolov, B.A. Brown, and V. Zelevinsky,
Phys. Rev. E {\bf 58}, 56 (1998).
\bibitem{cernaj} P. Cejnar, V. Zelevinsky, and V.V. Sokolov. Phys.
Rev. E {\bf 63}, 036127 (2001).
\bibitem{Bel1} S.T. Belyaev, Sov. Phys. JETP {\bf 12}, 968 (1961).
\bibitem{baranger65} M. Baranger, K. Kumar, Nucl. Phys {\bf 62}, 113 (1965).
\bibitem{sheikh02} J.A. Sheikh, P. Ring, E. Lopes, and R. Rossignoli,
Phys.Rev. C {\bf 66}, 044318 (2002).
\bibitem{vbz} A. Volya, B. A. Brown, and V. Zelevinsky,
Progr. Theor. Phys. Suppl. {\bf 146}, 636 (2002).
\bibitem{volya02} A. Volya, Phys. Rev. C {\bf 65}, 044311 (2002).
\bibitem{GDM} S.T. Belyaev and V.G. Zelevinsky, Yad. Fiz.
{\bf 16}, 1195 (1973).
\bibitem{Jap} V.G. Zelevinsky, Progr. Theor. Phys. Suppl.
{\bf 74-75}, 251 (1983).
\bibitem{jol} V. Zelevinsky and A. Volya, in {\sl Perspectives of
Nuclear Structure and Nuclear Reactions} (Dubna, 2002) p. 101.
\bibitem{wyom} V. Zelevinsky, in {\sl Mapping the Triangle}, AIP
Conf. Proc., vol. 638 (Melville, New York, 2002) p. 155.
\bibitem{Sol} V.G. Soloviev, Nucl. Phys. {\bf 9}, 655 (1958/59);
Mat. Fys. Dan. Vid. Selsk. {\bf 1}, No. 11 (1961).
\bibitem{Migd} A.B. Migdal, Nucl. Phys. {\bf 13}, 655 (1959).

\end{thebibliography}
\end{document}